\documentclass[manuscript]{aastex}

\usepackage{epsfig}
\usepackage{amssymb}
\usepackage{graphicx}

\shorttitle{The \emph{Herschel} DIGIT Survey of WTTSs}
\shortauthors{Cieza et al.}

\begin{document}


\title{The \emph{Herschel} DIGIT Survey of Weak-line T Tauri Stars:
implications for disk evolution and dissipation{\LARGE{$^{\star}$}}}


\author{Lucas A. Cieza\altaffilmark{1},
Johan Olofsson\altaffilmark{2},
Paul M. Harvey\altaffilmark{3},
Neal J. Evans~II\altaffilmark{3},
Joan Najita\altaffilmark{4},
Thomas Henning\altaffilmark{2},
Bruno Mer\'in\altaffilmark{5},
Armin Liebhart\altaffilmark{6},
Manuel G\"{u}del\altaffilmark{6},
Jean-Charles Augereau\altaffilmark{7}, and
Christophe Pinte\altaffilmark{7}
}
\email{lcieza@ifa.hawaii.edu}

\altaffiltext{}{$\star$ Herschel is an ESA space observatory with science
instruments provided by European-led Principal Investigator consortia and with
important participation from NASA. Based in part on observations made with the
CFHT, under program 11AH96}
\altaffiltext{1}{Institute for Astronomy, University of Hawaii at Manoa,
Honolulu, HI 96822. \emph{Sagan} Fellow, lcieza@ifa.hawaii.edu}
\altaffiltext{2}{Max Planck Institute f\"ur Astronomie, K\"onigstuhl 17, Heidelberg, Germany}
\altaffiltext{3}{Department of Astronomy, University of Texas at Austin,
2515 Speedway, Stop C1400, Austin, TX 78712-1205}
\altaffiltext{4}{National Optical Astronomy Observatory, 950 N. Cherry Avenue,
Tucson, AZ 86719, USA}
\altaffiltext{5}{Herschel Science Centre, European Space Astronomy Centre, ESA, P.O. Box 78, 28691 Villanueva de la Ca\~nada, Madrid, Spain}
\altaffiltext{6}{Department of Astronomy, Univ. of Vienna,
T\"{u}rkenschanzstr. 17, A-1180 Vienna, Austria}
\altaffiltext{7}{UJF-Grenoble 1/CNRS-INSU, Institut de Plan\'etologie et d'Astrophysique (IPAG) UMR 5274, BP 53, 38041 Grenoble cedex 9, France}

\begin{abstract}
As part of the ``Dust, Ice, and Gas In Time (DIGIT)" \emph{Herschel} Open Time
Key Program,
we present \emph{Herschel} photometry (at 70, 160, 250, 350 and 500 $\mu$m) of
31 Weak-Line T Tauri star (WTTS) candidates  in order to investigate the evolutionary status of their
circumstellar disks. Thirteen stars in our sample had circumstellar disks
previously  known from infrared observations at shorter  wavelengths, while
eighteen of them had no previous evidence for a disk.
We detect  a total of 15 disks  as all previously known disks are detected at
one or more \emph{Herschel} wavelengths and two additional disks
are identified for the first time.
The spectral energy distributions (SEDs) of our targets seem to trace the
dissipation of the  primordial disk and the transition to the debris disk
regime.
Seven of the 15 disks appear to be optically thick primordial disks, including
two objects with SEDs indistinguishable from
those of typical Classical T Tauri stars,   four  objects that have significant
deficit of excess emission at all IR wavelengths, and
one  ``pre-transitional" object with a known gap in the disk.
Despite their previous WTTS classification, we find that the seven targets in our sample with 
optically thick disks show evidence for accretion.  
The remaining eight disks have weaker IR excesses similar to those of optically thin debris disks.
Six of them are warm
and show significant 24 $\mu$m  \emph{Spitzer} excesses, while the last two are
 newly identified cold
debris-like disks  with photospheric 24 $\mu$m fluxes, but significant excess emission at
longer wavelengths.
The \emph{Herschel}  photometry also places strong constraints on the
non-detections, where systems
with $F_{70}$/$F_{70,\star}$ $\gtrsim$ 5--15  and 
$L_{disk}$/$L_{\star}$ $\gtrsim$ 10$^{-3}$--10$^{-4}$ can be ruled out.
We present preliminary models for both the optically thick and optically thin
disks  and discuss our results in
the context of  the evolution and dissipation of  circumstellar disks.

\end{abstract}


\section{Introduction}\label{intro}

Circumstellar disks are an unavoidable outcome of the star-formation process
and the
sites where planets form.  Understanding their evolution and dissipation is key
for
planet formation theory.
Early in their evolution, most optically visible pre-main-sequence (PMS) stars
are
surrounded by relatively massive ($\gtrsim$ 5 M$_{\mathrm{JUP}}$) disks that are still
accreting
material onto the stellar surface through magnetically channeled accretion
flows (Bouvier et al. 2007).  These objects are spectroscopically classified, 
 based on their strong H$\alpha$ emission,
as Classical T Tauri stars (CTTSs) 
or Herbig Ae/Be stars, the higher mass counterparts to CTTSs, if the stellar
mass is higher than
$\sim$2  M$_{\odot}$.
PMS stars without such evidence for ongoing  magnetospheric accretion are known
as weak-line T Tauri
stars (WTTSs).
Historically,  the dividing line between these two groups was  a  H$\alpha$
equivalent width (EW)  of
10 \AA. However, this  initial boundary was mostly driven by the sensitivity
limit of early
objective-prism surveys (e.g., Wilking et al. 1987) and has  been refined
multiple times in more
recent years (see Section~\ref{acc}).

Since they lack strong H$\alpha$ emission and (typically) near-IR excess,
WTTSs are less conspicuous than their accreting counterparts,
but can be identified as young PMS stars from  a number of youth indicators,
including  X-ray emission (Wichmann et al. 1996, 1997), Li I  (6707 \AA)
absorption
stronger than that of a Pleiades star of the same spectral type (Covino et al.
1997),
and/or their position  in the Hertzsprung--Russell diagram.
\emph{Spitzer} surveys of WTTSs (Padgett et al. 2006;  Cieza et al. 2007;
Wahhaj et al. 2011)
 have shown that the vast majority ($\gtrsim$ 80 $\%$) of WTTSs in nearby
molecular clouds  have lost their
 disks and that the few WTTSs that still retain circumstellar material  exhibit
a wide range of properties
 (e.g., SED morphology, disk inner radius, L$_{disk}$/L$_{\star}$)  that bridge
the gap  between the CTTS and debris disk regimes.
CTTSs  have primordial gas-rich disks and usually show optically thick
excess emission that extends all the way from the near-IR to the
 far-IR. Debris disks are gas poor  and are characterized by optically thin IR excess
emission that is mostly detectable in the far-IR and is believed to trace belts 
of second-generation dust produced by the collision of planetesimals (Wyatt 2008).
 The properties of WTTS disks overlap with those of both the
CTTS and debris disk population and may trace evolutionary stages between
the two.
Also, (sub)millimeter observations of WTTSs  show that there is a  very strong
correlation between the  lack of accretion and a very low disk mass
(Andrews $\&$ Williams, 2005, 2007; Cieza et al.  2010, 2012a).
WTTSs  are thus ideal laboratories to study disk evolution and, in particular,
their dissipation.

Circumstellar disks are complex systems that may evolve through a number of
physical  processes,  including  viscous accretion (Hartmann et al. 1998; Flock et al. 2011),
grain growth and dust settling (Brauer, Dullemond \& Henning 2008; Dullemond
$\&$ Dominik 2004),
photoevaporation (Alexander et al. 2006; Gorti et al. 2009; Owen et al. 2011),
and dynamical interactions with (sub)stellar companions (Artymowicz \& Lubow,
1994).
Planet formation through core accretion
(Lissauer $\&$ Stevenson, 2007) or gravitational instability (Durisen et al.
2007)  is also expected to play
a major role in disk evolution.
Furthermore, it is  increasingly clear that multiple evolutionary paths exist
and that not all disks evolve
in the same way or at the same rate (see  Williams $\&$ Cieza (2011) for a
recent
review on primordial circumstellar disks and their evolution).
Since WTTS disks have already undergone significant evolution,  their
incidence and properties
can provide valuable constraints on each one of the aforementioned  disk
processes and
on the transition between the primordial  and debris disk stages.

In this study, part of the ``Dust, Ice, and Gas In Time (DIGIT)"
\emph{Herschel} Open Time Key Program,
we present \emph{Herschel}  (Pilbratt et al. 2010) photometry (70 $\mu$m to 500
$\mu$m) of 31 WTTS candidates and  investigate
the properties and evolutionary status of their disks.
In \S 2 we describe our sample selection, the \emph{Herschel} observations, and
the data reduction.
In  \S 3 we describe the models we use to constrain the properties of both
optically thin and optically thick disks.
In \S 4 we discuss the nature of each one of our disks and
place our  results in the broader context of disk evolution.
Finally, a summary of our main results and conclusions is presented in \S 5.

\section{Observations and Data reduction}

\subsection{Target Selection}

We selected our sample from the literature of young stellar objects in nearby
(d $\lesssim$  200 pc) star-forming
regions and young stellar associations (age $\lesssim$ 10 Myrs).
The two  selection criteria were a previous WTTS classification and the availability
of \emph{Spitzer} data  (3.6 to 24 $\mu$m photometry and/or Infrared Spectrograph spectra) for disk identification purposes.
The WTTS classification criteria found in the literature are far from
homogenous,  but all
are based on the equivalent width or velocity dispersion  of the  H$\alpha$ line.
Despite their previous WTTS classification, we find evidence for accretion in 7 of our targets (see Section~\ref{acc} for
a detailed discussion).
Of our 31 targets,  8 are located in Taurus (Padgett et al. 2008), 5  in the 
$\eta$ and $\epsilon$ Chameleon associations (Sicilia-Aguilar et al. 2009; Brown et al. 2007),  
4 in the TW Hydra association (TWA, Low et al. 2005),
4 in Lupus, and 10 in Ophiuchus (Evans et al. 2007).
Thirteen of them had \emph{Spitzer}-identified disks, while the other eighteen
had no previous
evidence for a disk.
Twenty nine of our objects are K and M-type stars.
Two of them are G-type stars.
All of our targets are well-studied objects with multiple signatures of 
youth: they seem overluminous, show H$\alpha$ and X-ray emission, 
and/or Li I absorption stronger than that of a Pleiades star of the same 
spectral type.  They can thus be considered bona fide young stellar objects.
However, their individual ages are highly uncertain. In particular,  
individual interlopers as old as $\sim$100 Myr can not be ruled out 
(Padgett et al. 2006).
The most relevant stellar properties of our targets are listed in
Table~\ref{t:sample}, including multiplicity information
from adaptive optics imaging, aperture masking, and radial velocity 
observations.

\subsection{\emph{Herschel} photometry}

We obtained far-IR and submillimeter  wavelength photometry for  our sample of
WTTSs as part of the DIGIT  \emph{Herschel}  Key Program
using both the Photodetector Array Camera and Spectrometer (PACS, 70 and 160
$\mu$m, Poglitsch et al.  2010) and the Spectral and Photometric Imaging
Receiver
(SPIRE, 250, 350 and 500 $\mu$m, Griffin et al. 2010).  The PACS observations
were obtained
using the
Mini Scan Map  mode, while
the SPIRE observations were taken
using the  Small Scan Map mode.
Most of the targets were observed with PACS at two scan angles (70 and 100 deg)
for a better reduction of the
1/f noise. Due to observing time constraints in the DIGIT program, isolated objects free of significant cloud emission (e.g.,   TWA and
$\eta$ Cha targets) were observed with a single PACS scan.
The identification number, duration, and date of the PACS and SPIRE
observations are listed in Table~\ref{obs_ids}.

Both the PACS and SPIRE data were processed using HIPE (Herschel Interactive
Processing
Environment; Ott et al. 2010) version 7.1. 
We utilized a standard high-pass filtering script for PACS, appropriate for point-source observations,  
and a standard baseline-subtraction pipeline for SPIRE.  
When applicable, the two separate PACS OBSID�s for a given object were processed together.  
The portions of scan legs where the telescope was stopped or not scanning at a uniform velocity were not used.
The PACS beam sizes at 70 and 160
$\mu$m are 5.5$\arcsec$ and 11$\arcsec$,
and the data were resampled to 1$\arcsec$ and 2$\arcsec$ pixel images,
respectively. SPIRE beam sizes at 250, 350 and
500 $\mu$m are  18.1$\arcsec$, 25.2$\arcsec$ and 36.6$\arcsec$,  and  the
images were resampled to 6$\arcsec$,
10$\arcsec$, and 14$\arcsec$ pixel images,
respectively.
In all cases, the pointing accuracy of the telescope ($\lesssim$ 2$\arcsec$; Eiroa et al. 2012)
is significantly smaller than the beam size. 
As the next step in the data reduction, we ran the source-finding and
PSF-fitting program
\emph{c2dphot} on the image produced by HIPE.
The \emph{c2dphot} software was used for the entire {\it c2d Spitzer} Legacy
Project (Evans et al. 2003; 2009)
and is based on the DOPHOT program (Schechter et al. 1993).
This software uses an adjustable square box to fit sources with a (possibly)
inclined plane background plus a  point spread function (PSF).   
In the case of the \emph{Herschel} data, we used a 9-pixel box.  Because of this approach, the derived PSF-fit
flux is relatively insensitive to the level or gradient in the background
as long as the background is not highly structured on the scale of the
PSF.

\begin{figure}
\includegraphics[width=12cm, angle=270, trim = 0mm 0mm 0mm 0mm, clip]{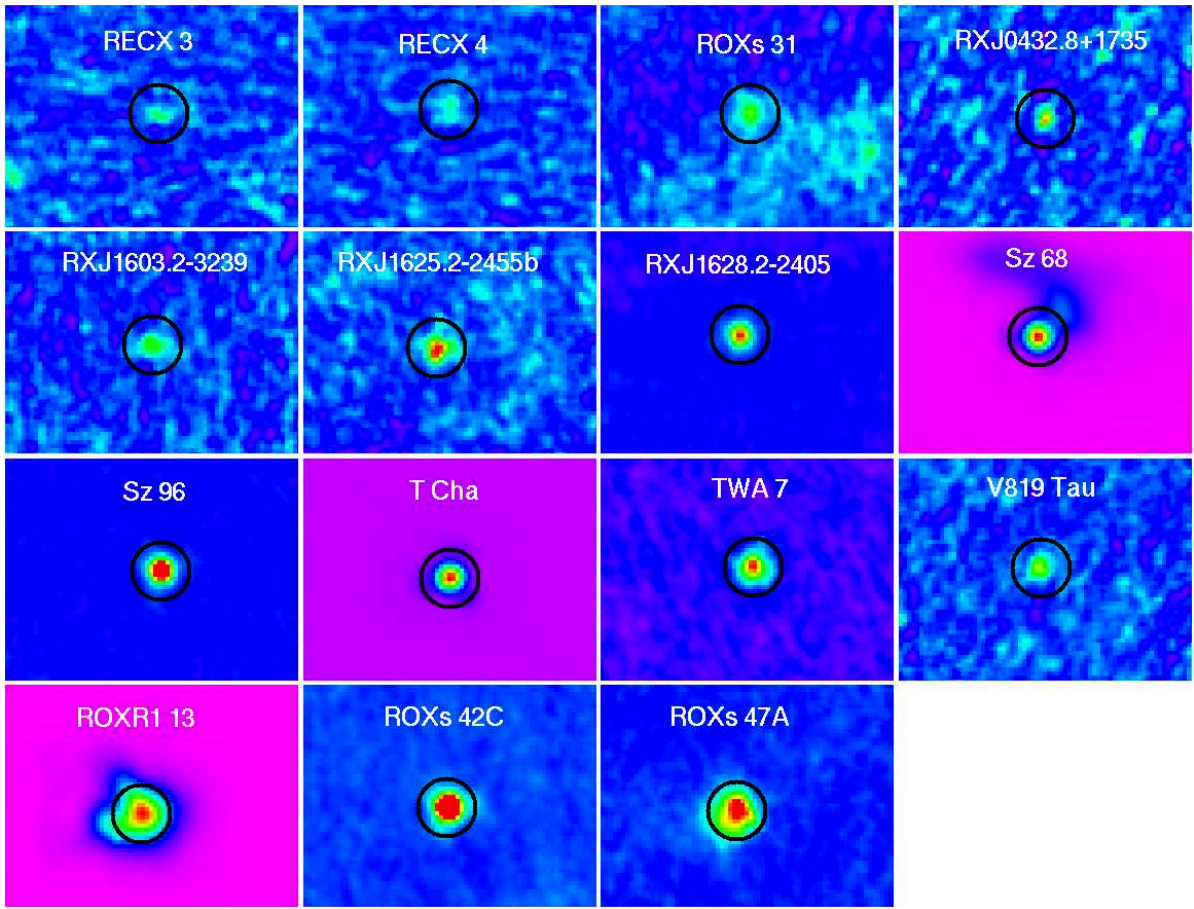}
\caption{PACS images of the 15 objects detected at 70 $\mu$m. The black circles
are 7$\arcsec$ in
radius and indicate the nominal positions of the targets. North is up. East is to the left.}
\label{DET-70}
\end{figure}

\begin{figure}
\includegraphics[width=10cm, angle=270, trim = 0mm 0mm 0mm 0mm, clip]{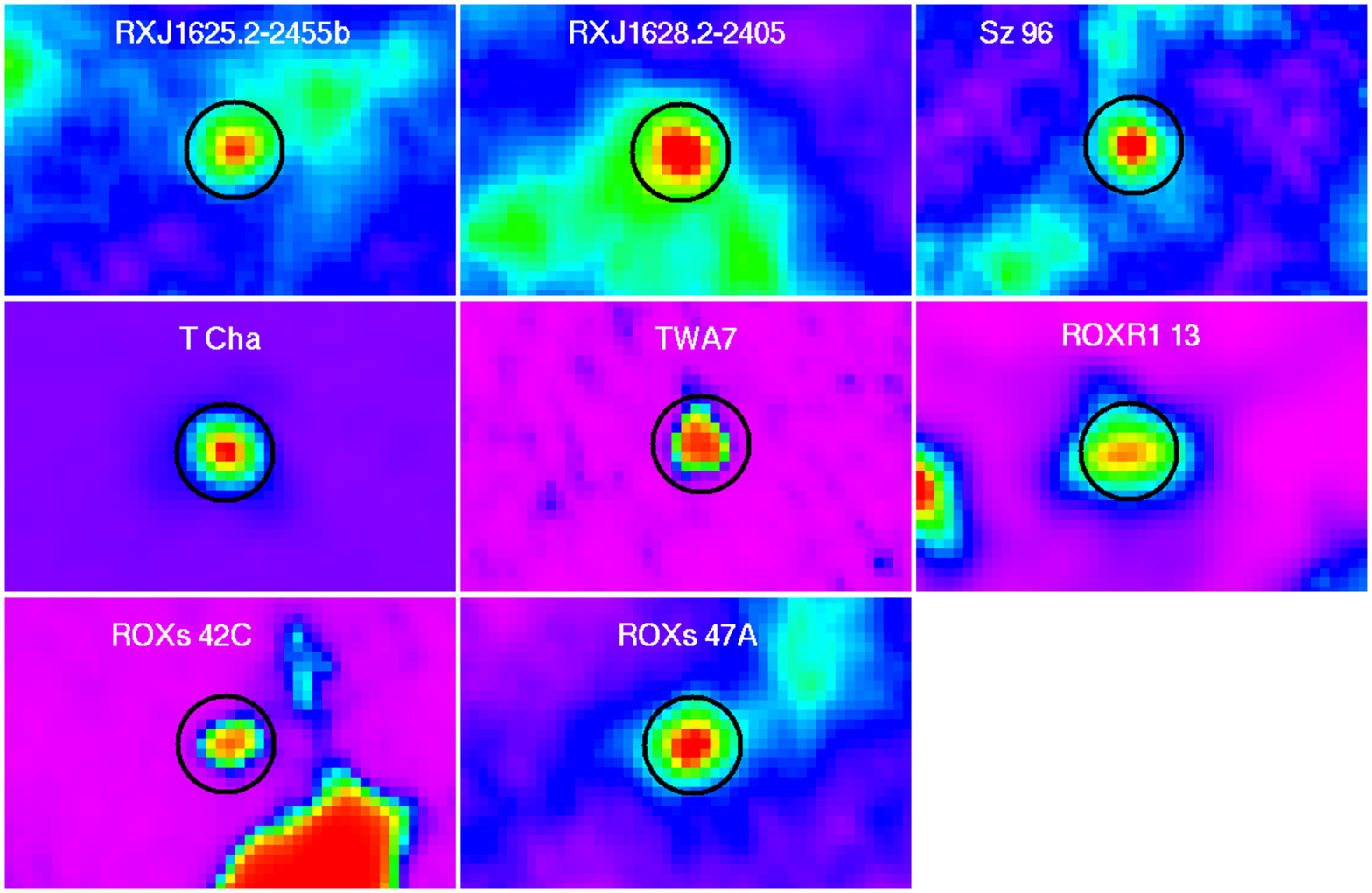}
\caption{PACS images of the 8 objects detected at 160 $\mu$m. The black circles
are 14$\arcsec$ in
radius and indicate the nominal positions of the targets.  North is up. East is to the left.}
\label{DET-160}
\end{figure}

\begin{figure}
\includegraphics[width=9cm, angle=270, trim = 0mm 0mm 0mm 0mm, clip]{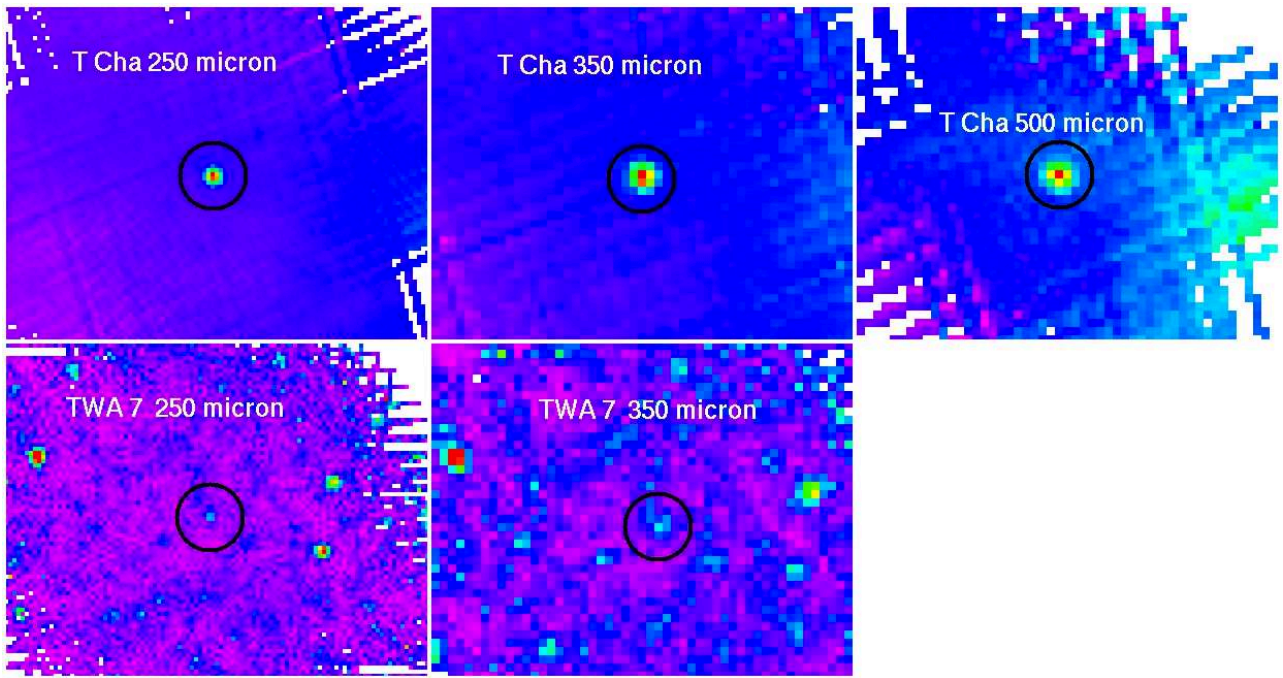}
\caption{SPIRE  images (250, 350, and/or 500 $\mu$m)  of the 2 targets detected
at submillimeter wavelengths.   The black circles are 50$\arcsec$ in
radius  and indicate the nominal positions of the targets. North is up. East is to the left.}
\label{DET-SPIRE}
\end{figure}

We first ran  \emph{c2dphot} in its standard mode, which finds peaks above the
background and fits a  PSF to the local maxima.   We used the empirical
PSF's posted by the NASA Herschel Science
Center\footnote{https://nhscsci.ipac.caltech.edu/sc/index.php/}.
The \emph{c2dphot}  algorithm detected (with S/N $\gtrsim$ 5) fifteen of our targets at 70 $\mu$m, 
the images of which are shown in Figure~\ref{DET-70}.
Thirteen of these targets had previously known 24 $\mu$m excesses.
RXJ1625.2-2455b and ROXs 31 are newly identified disks.
Similarly,  8 of our  targets  were detected at 160 $\mu$m (see
Figure~\ref{DET-160}).
At SPIRE wavelengths, we detected only two objects: T~Cha and TWA 7.  T~Cha was
detected in
all three SPIRE bands, while TWA 7 is seen only at 250 and 350 $\mu$m.
Their SPIRE images are shown in Figure~\ref{DET-SPIRE}.
Most of the detections are consistent with point sources.
Nevertheless,  the 70 $\mu$m images of  ROXR1~13 and ROXs~47A  and  the
160 $\mu$m images of  RXJ1628.2-2405, Sz~96, ROXR1~13,
ROXs~42C,  and ROXs 47A  seem to suffer from different levels of
contamination by cloud emission because \emph{c2dphot} finds ``extended
sources" at the location of the targets.
The images of Sz 68 show an extended source or nebulosity north of the target,
which does not seem
to affect the 70 $\mu$m detection,  but  completely dominates the emission at
longer wavelengths.
In these cases, we re-ran  \emph{c2dphot} in a mode that forces the
program to fit a PSF at the  position of the target.
The use of this PSF-fitting  mode should mitigate the flux contamination, but
it is quite possible
that we are still overestimating the far-IR fluxes of the  sources listed
above.
To be conservative, we treat the fluxes suspected of contamination  as upper
limits
when modeling the disks is Section~\ref{modeling}.

Since the uncertainty images produced by HIPE 7.1 are not reliable, especially for PACS data,
we  follow Harvey et al. (2012a; 2012b) and estimate the photometric uncertainties  
by using \emph{c2dphot}  to estimate the noise at
eight fixed positions around the target, but within the high-coverage area.
For each one of  the undetected objects,  a similar grid was run to estimate
upper limits from
the noise at and around the location of each target.
The resulting \emph{Herschel} photometry and 3$-\sigma$ upper limits
for our 31 objects are listed in Table~\ref{t:photometry}.
In low background regions, we  find 3-$\sigma$ upper  limits of  the order of
$\sim$3,  10,  20,  20, and 30 mJy  at 70, 160, 250, 350, and 500 $\mu$m,
in agreement with the prediction of the \emph{Herschel} Observation Planning
Tool.
However, these limits can be an order of magnitude higher in regions of strong
background emission,
especially at  longer wavelengths.

\subsection{High-resolution optical spectroscopy}

Seven of our targets (T Cha, Sz 68, Sz 96,  ROXR1 13, RXJ1628.2-2405, ROXs 42C,
and ROXs 47)
show  significant excesses at the wavelengths sampled by  \emph{Spitzer}'s
Infrared Array Camera (IRAC, 3.6 -  8.0 $\mu$m).
Since there is a \emph{very} strong correlation between the presence of
near-IR excess and accretion indicators (Hartigan et al.1995), these objects
can be suspected to be weak accretors  despite their WTTS classification.

In order to firmly establish their accretion status, we obtained Echelle
spectroscopy   for
6 of them using
the ESPaDonS Echelle spectrograph on the 3.5-meter Canada-France-Hawaii
Telescope (CFHT)
at Mauna Kea Observatory in Hawaii.
T Cha was excluded because it is far south and not visible from Mauna Kea.
The observations were performed in service mode
between June 06  and  July 05, 2011 (the 2011A semester), under program 11AH96.
The spectra were obtained in the standard ``star+sky" mode,
which delivers the complete optical spectra between 3500 \AA \
and 10500 \AA \  at a resolution of 68,000, or 4.4 km/s.
For each object, we obtained  a set of 2 spectra with exposures times
ranging from 20 to 30 minutes each, depending on the brightness
of the target. The data were reduced through the standard CFHT pipeline
Upena, which is based on the  reduction package
Libre-ESpRIT\footnote{ \tt
http://www.cfht.hawaii.edu/Instruments/Spectroscopy/Espadons/Espadons\_esprit.html}.
In the following section,  we use these data to investigate the accretion
properties of the 6 CFHT targets.

\section{Results}

\subsection{Accreting weak-line T Tauri stars?}\label{acc}

Before investigating the circumstellar properties of our objects, we review
the accretion status of our  targets.
Nominally, CTTSs are accreting circumstellar material and WTTSs are
non-accreting objects.
However, since most WTTSs show H$\alpha$ emission from chromospheric activity,
establishing
an exact observational boundary between WTTSs and CTTSs is problematic.
As mentioned in Section~\ref{intro}, the dividing line between CTTSs and WTTSs
was  initially set at a
 H$\alpha$ equivalent width (EW)  of  10 \AA~and was mostly
driven by the sensitivity limit of early  objective-prism surveys (e.g.,
Wilking et al. 1987). This boundary  has
 more recently been refined to take into account different continuum levels of
stars of different spectra types.
Mart\'in (1998) suggested the following boundary  dependence with spectral
type:
5 \AA~for spectral types earlier than M0, 10 \AA~for M0 to M2 stars, and 20
\AA~for later spectral
types. White \& Basri (2003) extended this dependence to hotter and cooler
stars
and suggested 3 \AA~for spectral types earlier than K0 and 40 \AA~for objects
later than M5.
Nevertheless, the width  ($\Delta V$, measured at 10$\%$ of the peak emission)
and shape of the H$\alpha$ profile obtained from high-resolution spectroscopy
are now known to be more sensitive accretion diagnostics than H$\alpha$ EWs
derived from
low-resolution data (Natta et al. 2004).
Non-accreting objects show narrow ($\Delta V \lesssim$ 230-270 km/s) and
symmetric line
profiles of chromospheric origin, while accreting objects present broad ($\Delta
V \gtrsim$  230-270 km/s)
and asymmetric profiles produced by high-velocity magnetospheric accretion
columns
(Jayawardhana et al. 2003; White \& Basri 2003).
Regardless of the exact boundary used, magnetospheric accretion is
highly variable (Jayawardhana et al. 2006; Nguyen et al. 2009) and can be
sporadic.
Thus, borderline  objects can move
back and forth between the CTTS and WTTS classification (e.g., Murphy et al.
2011).

\begin{figure}
\includegraphics[width=15cm, trim = 0mm 0mm 0mm 0mm, clip]{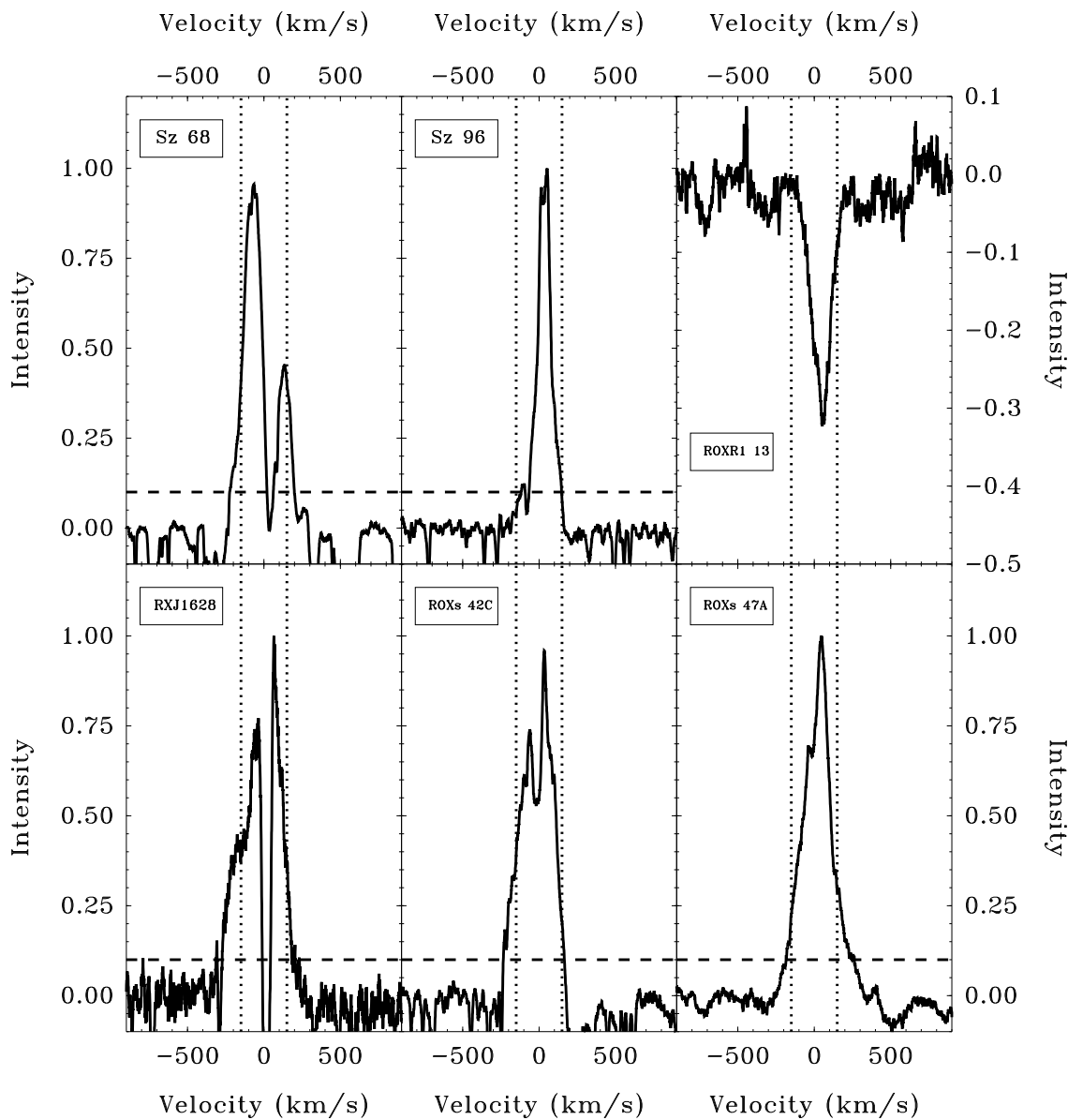}
\caption{The H$\alpha$ velocity profiles of the 6 optically thick disks
observable from Mauna Kea.
T Cha is excluded as it has a declination of $-$79 deg.
The vertical dotted lines correspond to  a velocity width of 300 km/s.
With the exception of  ROXR1 13, all objects show
$\Delta$V $\gtrsim$ 270 km/s and asymmetric emission lines, indicating
accretion.
ROXR1 13 has been previously shown to have a very variable H$\alpha$ profile,
ranging from pure absorption to broad emission
(Jensen et al. 2009).
}
\label{f:halpha}
\end{figure}

Figure~\ref{f:halpha} shows the H$\alpha$ velocity profiles of the 6 targets we
have observed with the CFHT. 
We find evidence for weak but detectable accretion ($\Delta V$ $\gtrsim$ 270
km/s)
in all of the CFHT targets except for ROXR1~13.
However, ROXR1 13 has been previously shown by Jensen et al.  (2009) to have a
very variable
H$\alpha$ profile, ranging from pure absorption to broad emission ($\Delta V
\gtrsim$  400 km/s).
We also collected $\Delta V$ values from the literature for most of the other
targets
(see Table~\ref{t:sample}). Except for T Cha, all  the literature values are
consistent with chromospheric
activity ($\Delta$V $\lesssim$ 230 km/s).
Similar to  ROXR1~13, T~Cha is also known to have a very variable H$\alpha$
profile suggesting sporadic
accretion (Schisano et al. 2009).

\subsection{Primordial disks, debris disks, and diskless stars}\label{class}

\begin{figure}
\includegraphics[width=14cm, trim = 0mm 0mm 0mm 0mm,clip]{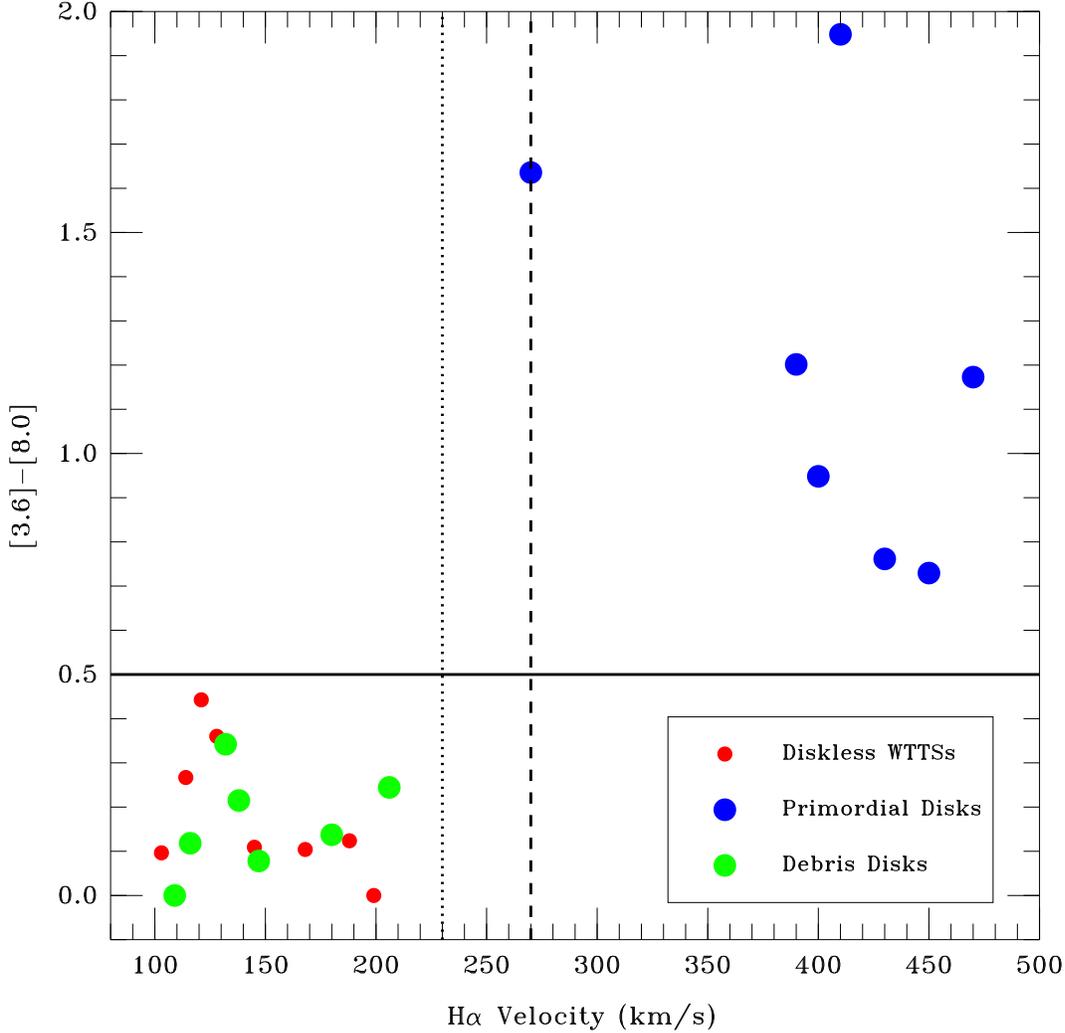}
\caption{
The  [3.6]--[8.0] color as a function of the velocity width of the H$\alpha$
line ($\Delta$V)  for all the objects for
which  $\Delta$V measurements are available.
An IRAC color of 0.0 has been assigned to TWA 7 and TWA 10 as there
 is no IRAC data available for them,   but their IRS spectra show no excess
shortward of 10 $\mu$m. The dotted and dashed vertical lines correspond to the
accretion detectability boundaries adopted by Jayawardhana et al.  (2003) and
White $\&$ Basri (2003),
respectively.
The horizontal line divides the objects with and without IRAC excesses.
We find two distinct populations in our sample:  1) accreting objects  with
IRAC excesses, consistent with optically thick primordial disks (blue dots),  and
2) non-accreting objects without IRAC excesses, which are likely to be
either diskless stars (red dots) or debris disks detectable at $\lambda$ 
$\gtrsim$ 20 $\mu$m (green dots). 
}
\label{H-vs-I1I4}
\end{figure}

In Figure~\ref{H-vs-I1I4}, we plot the [3.6]--[8.0] colors of our sample as a
function of $\Delta V$ of the H$\alpha$ line.
We find that all the targets with 8.0  $\mu$m excess show detectable levels of
accretion, while
none of the targets with photospheric IRAC colors do so.
Accreting objects with near-IR excess can safely be considered gas-rich
primordial disks.
Non-accreting objects without  near-, mid-, or far-IR excesses can be regarded
as diskless stars within
the limits of the survey.
However, non-accreting objects with  near-IR photospheric colors but
significant mid- or far-IR excesses
are more difficult to define.
They are likely to be the ``remnants"
of primordial disks that have dissipated through accretion and
photoevaporation.
They contain the material that has not yet  fallen into the star, 
formed large solid bodies,  
been evaporated,  or been blown away by radiation pressure.
Whether these objects share \emph{all} of the properties of older debris disks
(very gas-poor systems where the opacity is dominated by second-generation dust
produced by the collision of planetesimals) remains to be established.
Nevertheless,  we refer to them as debris disks throughout the paper because
they do show many of
the key properties of  debris disk systems: L$_{disk}$/L$_{\star}$ $<$ 0.1,
M$_{dust}$ $<$ 1 M$_{\oplus}$,
and lack of detectable accretion (Wyatt 2008).
Non-accreting objects with \emph{Herschel} excesses are hence
labeled  ``debris disks" in Figure~\ref{H-vs-I1I4}.

\subsection{Full SEDs}

\begin{figure}
\includegraphics[width=9cm, trim = 0mm 0mm 0mm 0mm, clip]{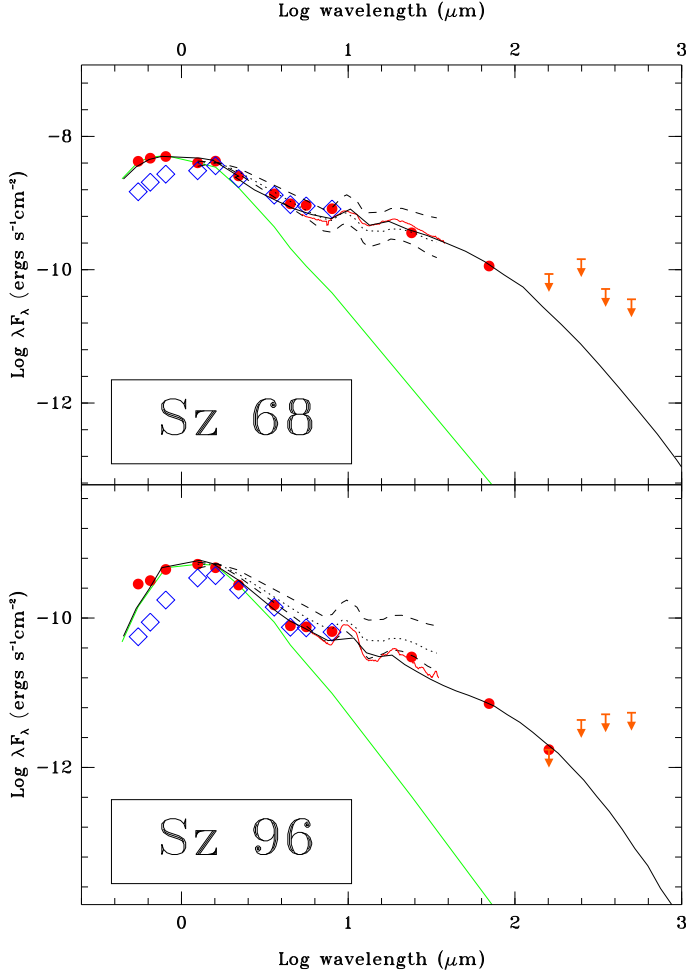}
\caption{The SEDs of Sz 68 and Sz 96, two targets indistinguishable from
typical CTTSs.
The filled circles are detections while the arrows represent 3--$\sigma$
limits.
Both symbols are shown when the objects is detected, but the flux is suspected
of being
contaminated by extended emission.
The open squares correspond to the observed optical and near-IR fluxes before
being corrected for extinction using the
R-I color excess (A$_V$=  4.76$\times$E[R-I]) and the extinction curve provided
by the Asiago database of photometric systems
(Fiorucci $\&$ Munari 2003). The red line is the \emph{Spitzer}-IRS spectrum.
The solid green line represents the stellar photosphere normalized to the
extinction-corrected J-band. The dotted line corresponds to the median mid-IR
SED of
K5-M2 CTTSs calculated by Furlan et al. (2006). The dashed lines are the
quartiles.
The solid black lines correspond to the models discussed in
Section~\ref{model-borderline}.
}
\label{f:border-sed}
\end{figure}

\begin{figure}
\includegraphics[width=15cm, trim = 0mm 0mm 0mm 0mm,clip]{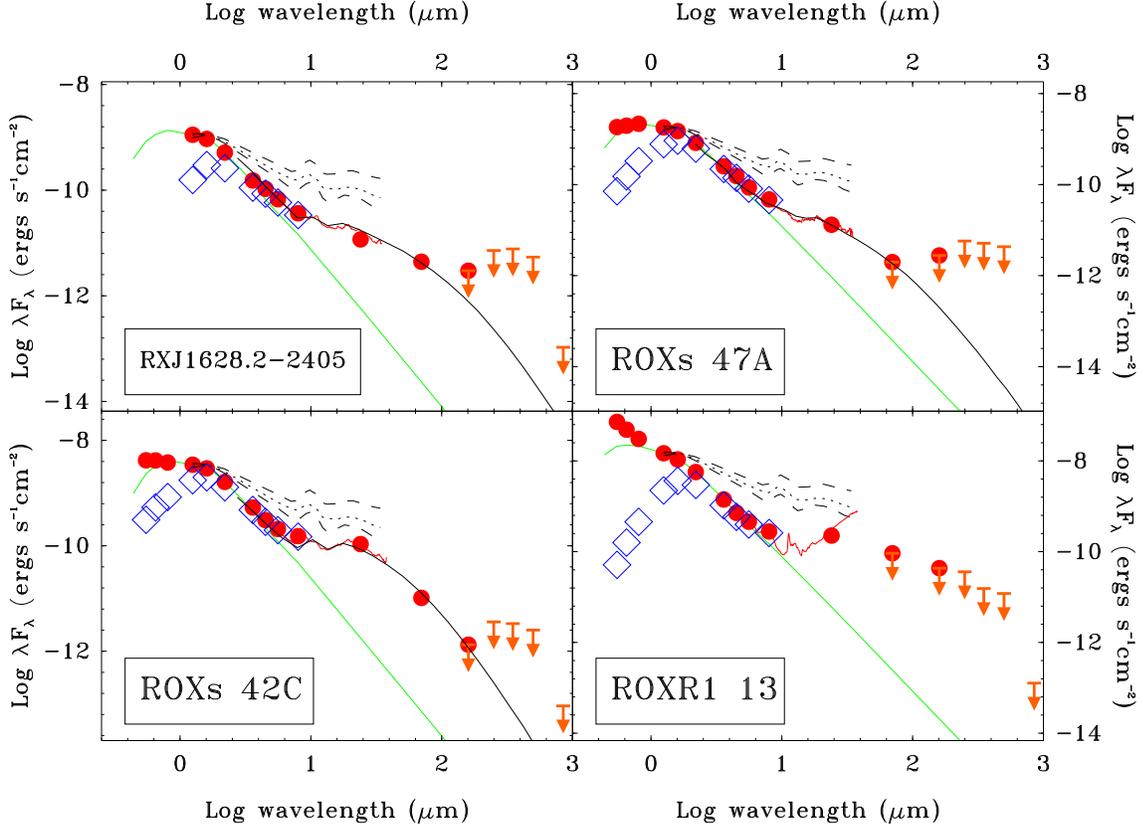}
\caption{
The SEDs of the 4 accreting targets with reduced levels of near-IR excess with
respect to typical
CTTSs. These systems can be considered to be ``evolved" primordial disks.
The symbols are  as in Figure~\ref{f:border-sed},  but in this case
the extinction has been estimated from the  J-K color excess (A$_V$=
5.87$\times$E[J-K]).
Both filled circles and arrows are shown when the objects is detected, but the
flux is suspected of being
contaminated by extended emission.
The solid black lines correspond to the models discussed in
Section~\ref{model-evolved}.
}
\label{f:primordial-sed}
\end{figure}

\begin{figure}
\includegraphics[width=15cm, trim = 0mm 0mm 0mm 0mm, clip]{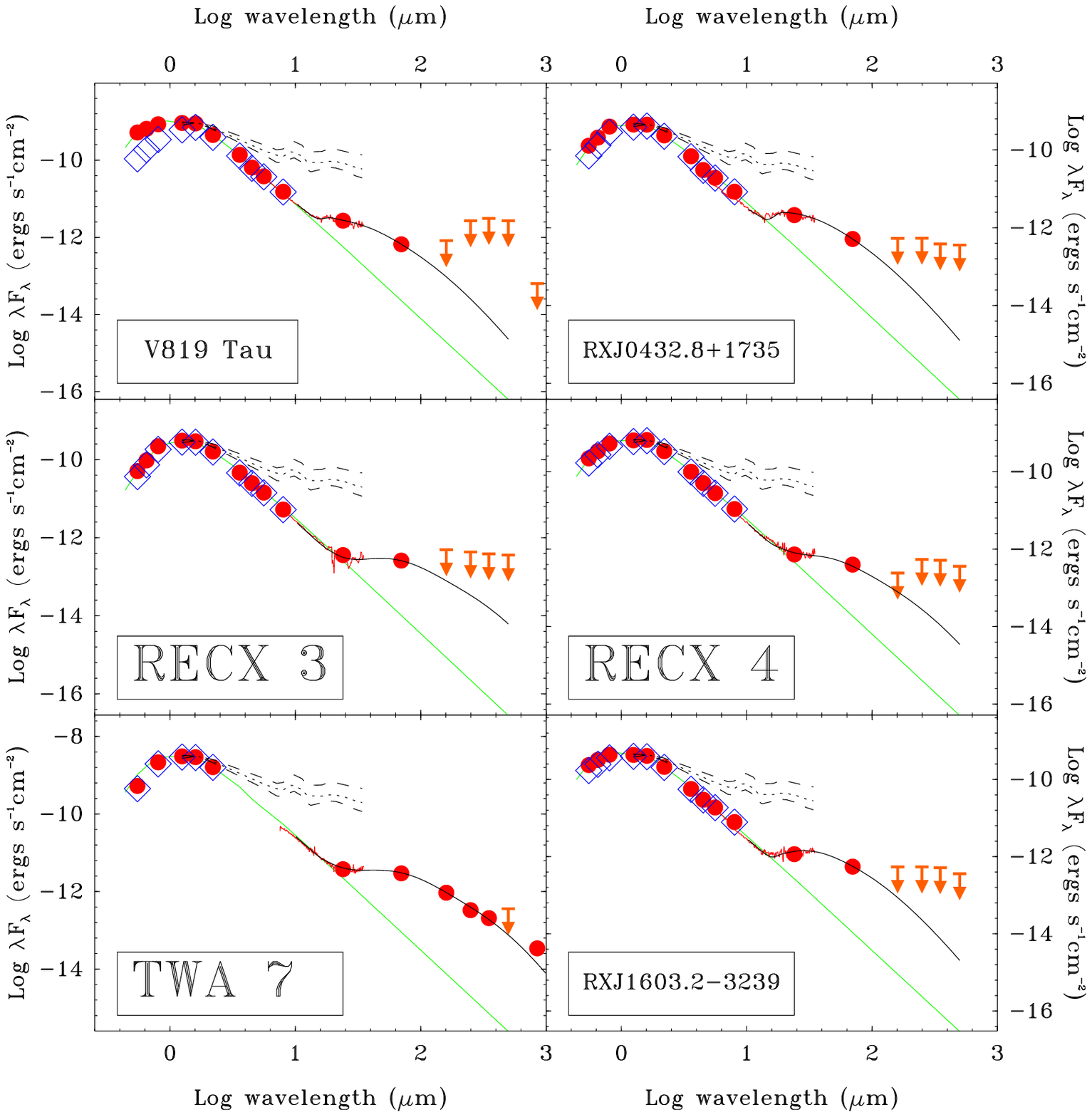}
\caption{The SEDs of the 6 ``warm" debris disks in  our sample. The symbols are
 as in
Figure~\ref{f:border-sed}. The extinction has been estimated from the  J-K
color excess (A$_V$=  5.87$\times$E[J-K]).
The solid black lines correspond to the models discussed in
Section~\ref{model-warm}.
}
\label{f:warm-sed}
\end{figure}

\begin{figure}
\includegraphics[width=12cm, trim = 0mm 00mm 0mm 00mm,clip]{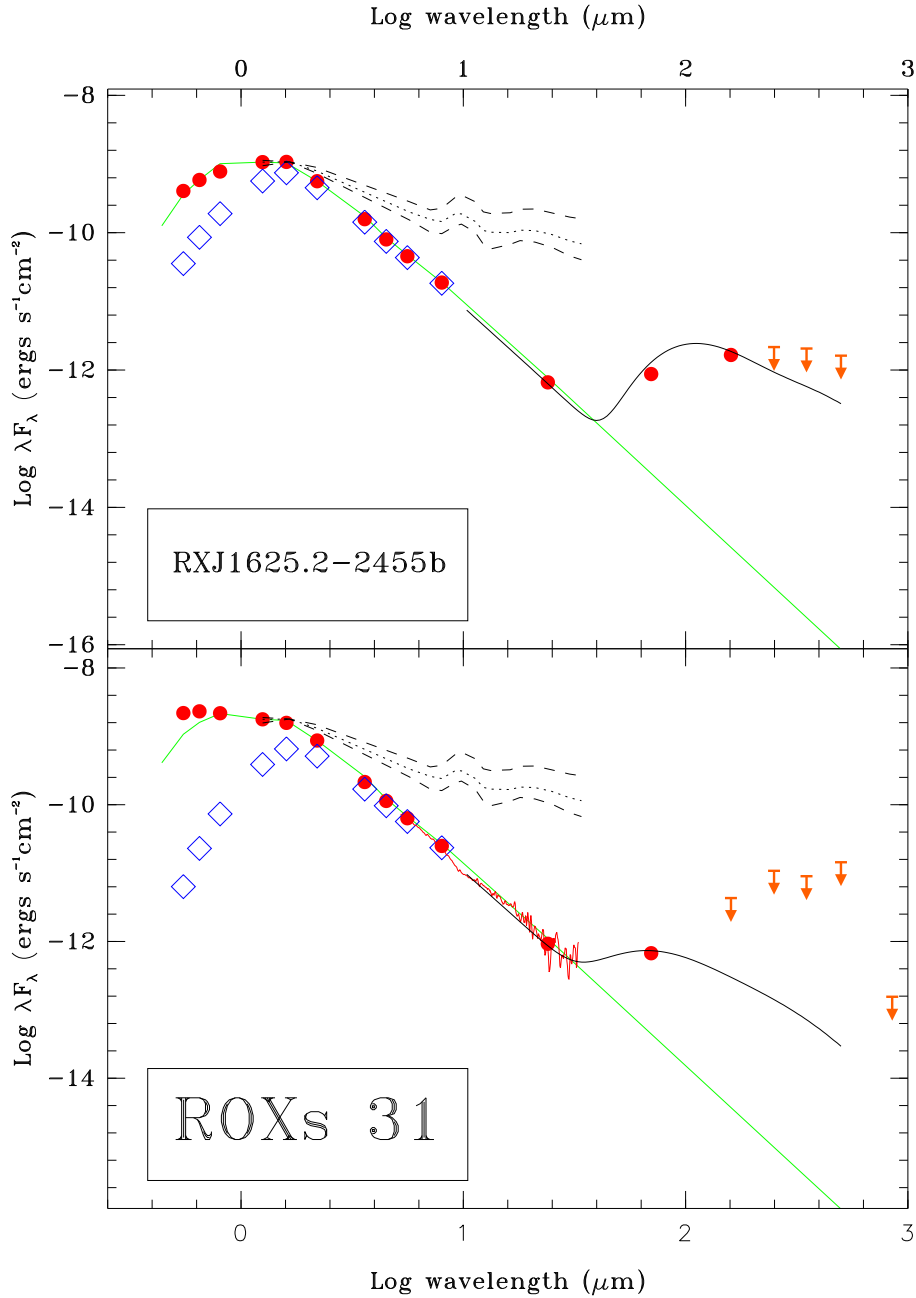}
\caption{The SEDs of the 2 ``cold" debris disks in  our sample. The symbols are
 as in
Figure~\ref{f:border-sed}. The extinction has been estimated from the  J-K
color excess (A$_V$=  5.87$\times$E[J-K]).}
The solid black lines correspond to the models discussed in
Section~\ref{model-cold}.
\label{f:cold-sed}
\end{figure}

We collected optical to (sub)millimeter photometry of all our targets from
the literature (see Table~\ref{t:photo-rest}) and constructed
the full SEDs of the 15 objects with IR excesses.
With the exception of  RXJ1625.202455b, we also found Infrared Spectrograph
(IRS) data for all of them
in the \emph{Spitzer} Archive.
We find a huge diversity in SED morphologies in our sample.
Two of the targets, Sz  68 and Sz 96, have SEDs that are  indistinguishable
from
those of typical CTTSs (see Figure~\ref{f:border-sed}).
Four  other objects (ROXR1 13, RXJ168.2-2405, ROXs 42C, and ROXs 47A) have
detectable IRAC excesses that
fall well below the median SEDs of CTTSs (see Figure~\ref{f:primordial-sed}).
The 6 objects  in these first two groups are accreting  primordial disks.
We also find 6 objects (V819 Tau, RXJ0432.8+1735, REXC 3, RECX 4, TWA 7 and
RXJ1603.2-3239)
with photospheric IRAC fluxes but significant excesses at $\lambda \gtrsim$  24
$\mu$m.
These objects are non-accreting and we consider them to be ``warm" debris disks
(see Figure~\ref{f:warm-sed}).
There are two targets (RXJ1625.2-2455b, ROXs 31) with photospheric 24 $\mu$m
fluxes
 that have significant \emph{Herschel} excesses (see Figure~\ref{f:cold-sed}).
We refer to these non-accreting objects as ``cold" debris disks.
Finally, there is T Cha, a pre-transition disk with strong near-IR excess, but
showing a  pronounced ``dip"
in the mid-IR that indicates the presence of a wide gap. The SED of T~Cha has
already been fully modeled
by Olofsson et al. (2011) and Cieza et al. (2011).
The  photometry  used to construct the SEDs corresponds to many different 
epochs,  and young stellar objects can be highly variable in the optical and 
the near-IR (Herbst et al.  1982, Carpenter et al. 2001). This may explain 
the poor fit to the stellar photospheres in some of the objects (e.g.,  
ROXR1 13 and ROXs 31).

\subsection{SED modeling}\label{modeling}

\subsubsection{Primordial disk models}\label{prim-models}

As discussed in Section~\ref{class},
 all of our targets with IRAC (3.6 to 8.0 $\mu$m) excesses
seem to be  accreting primordial disks.  To model the SEDs of these objects
(except for T~Cha),
we use the MCFOST radiative transfer code (Pinte et al. 2009).
We describe the structure of the primordial disks with the following parameters:
 the inner and outer radii of the dust disk (R$_{\mathrm{in}}$ and  R$_{\mathrm{out}}$, respectively), the index
 $\alpha$ for the surface density profile
  ($\Sigma$ (r) = $\Sigma_{50}$ (r/50 AU)$^{\alpha}$ )
  and a dust disk scale height $H(r)$
 assuming a vertical gaussian distribution (exp[$-$z$^2$/2H(r)$^2$]).
 The disk's flaring
is described by a power-law determining the scale height as a
function of radius ($H(r)$ = H$_{50}$(r/50 AU)$^{\beta}$). The dust content is
described by a differential power-law for the grain size distribution
(d$n$(a) $\propto$  $a^{p}$d$a$), between the minimum ($a_{\mathrm{min}}$) and maximum
($a_{\mathrm{max}}$) grain sizes.
Dust stratification (i.e., small grains in the surface layer and large grains
in
the midplane) is needed to simultaneously explain the silicate features
and millimeter wavelength  slopes of many protoplanetary disks.
Following Pinte et al. (2008), we hence adopt a parametric law to describe dust
stratification,
according to which scale height is a function grain size:
$H(a) = H(a_{\mathrm{min}})(\frac{a}{a_{\mathrm{min}}})^{-\xi}$.
With this prescription, $\xi$ = 0 corresponds to  a disk where
the gas and dust are perfectly mixed (i.e., no dust stratification).
For the grain composition, we use amorphous silicate grains with olivine
stoichometry (Dorschner et al. 1995).

SED modeling is known to be highly degenerate without spatially resolved
observations.
We thus fix several of the parameters and try to constrain others.
In particular,  we adopt $p = -$3.5, consistent with grains in collisional
equilibrium,
        $a_{\mathrm{max}}$ = 3.9 mm, corresponding to 3 times the size of the longest
SED point available,
and $a_{\mathrm{min}}$ = 0.1 $\mu$m, a size sufficiently small to have a little effect
on the SED.
We also set R$_{\mathrm{out}}$ = 100 AU for single stars and $\alpha$ = $-$1 as these
parameters are particularly difficult to constrain without spatially resolved
observations (Andrews et al. 2010).
For close binary systems, by which we mean that the secondary is close
enough to affect the disk, 
 R$_{\mathrm{out}}$ is set to value smaller than the \emph{projected} separation.
Finally, we adopt $\xi$ =  0.0, except for systems where
some dust stratification is needed to explain a particularly
strong silicate feature, indicating that  small grains
dominate the regions probed by the mid-IR data (e.g.,  the surface layer of the disk).  
We are thus left with only 4  truly free parameters:  $\Sigma_{\mathrm{50}}$,  R$_{\mathrm{in}}$,
 H$_{\mathrm{50}}$, and $\beta$.
These 4 parameters allow us to address 3  key disk evolution questions, namely
1) what are the surface densities and masses of these disks?  
2) do they have inner holes? and 
3) do they present  evidence for dust settling (e.g., flat structures and/or dust stratification)?

The disk mass (M$_{disk}$) is not a free parameter, but we estimate this
important quantity by integrating the surface density profile of the dust
over radius
and assuming a gas to dust mass ratio of 100.
M$_{disk}$ is very dependent on  the values adopted for a$_{\mathrm{max}}$ and the gas
to dust mass
ratio, which are highly uncertain  and expected to evolve with time due to disk
evolution processes
such as  photoevaporation, grain growth, and planet formation.
We thus emphasize  that the quantity constrained by our models is the total
mass of dust  with grain
sizes from $a_{\mathrm{min}}$ to $a_{\mathrm{max}}$ that is located between R$_{\mathrm{in}}$ and
R$_{\mathrm{out}}$; thus, the M$_{disk}$ values derived should be taken with caution.
In all cases, we fix the inclination of  the disk to 60 deg. 
We note however that,  unless the disk is very  close to edge-on ($i$ $\gtrsim$ 85 deg), 
the inclination has no discernible effect on the emerging SED. 
The default parameters for our primordial disks models are listed in
Table~\ref{t:pri-fixed}.
As mentioned above, the default $\xi$ value is allowed to change, but only if
needed to better fit the silicate feature. Similarly, R$_{\mathrm{out}}$ is allowed to change to a
lower
value, but only if the target is a close binary system.
As  illustrated by  our modeling work on T Cha (Cieza et al. 2011),  
unique disk parameters
cannot be obtained from the SED alone, even if the SED is very well sampled.
For each of the remaining optically thick disks, we therefore aim to find one
model that fits the available data  relatively well.
The ``best fit" models were obtained by trial and error, varying one
parameter at the time in search of a $\chi^2$ minimum.
For fitting purposes, the IRS spectra were binned and 
each bin treated as a photometry point. Upper limits were not used, except
to discard  models with fluxes exceeding those limits. 
The results thus must be considered to be highly preliminary.
However, even the preliminary  models presented herein are  expected to capture
some of the key properties  of the disks (e.g., the presence of a hole or dust
settling)  and
can be used as the starting point for future, more detailed models.
For ROXs 47A and ROXs 42C,  we were unable to satisfactorily reproduce the SEDs
using the parameterization discussed above. Instead, we present
 2-zone disk models with different sets of parameters for the inner
and outer disks (see Section~\ref{model-evolved}).
Similar 2-zone structures have already been invoked to simultaneously
fit the SEDs and resolved submillimeter images of
several other protoplanetary disks (e.g., Andrews et al. 2011; Cieza et al.
2012b; Mathews et al. 2012) and seem to be
fairly common.
The parameters for the best-fit models  of  the 6  primordial disks  are listed
in Table~\ref{t:pri-fit}.
For the disk parameters of the T Cha models, see Table  3 in Cieza et al.
(2011).

\subsubsection{Debris disk models}\label{thin}

To model the objects that we have classified as cold and warm debris disks, we used
the DEBRA code presented in Olofsson et al. (2012). Given the overall small
signal-to-noise ratio of the \emph{Spitzer}-IRS spectra, and the lack of any
strong emission features associated with silicate dust grains, we used only one
dust composition of astronomical silicates (optical constants from Draine
2003). We used the distribution of hollow spheres scattering theory (Min et al. 2005) 
to compute absorption efficiencies for sizes between $a_{\mathrm{min}}=0.1$\,$\mu$m and
$a_{\mathrm{max}}=1$\,mm. We checked that for each source, the adopted
$a_{\mathrm{min}}$ value does not violate the blow-out size. We computed the
unitless $\beta_{\mathrm{rp}}$ ratio between radiation pressure and
gravitational forces, for each star.  The $\beta_{\mathrm{rp}}$   value as a
function of grain size $a$ are shown in Figure\,\ref{fig:beta}. The threshold
value for a rapid evacuation of dust grains is $\beta_{\mathrm{rp}} > 0.5$ (see
Krivov 2010), indicating that radiation pressure has a very limited effect on
the observed debris disks in our sample. To limit the number of parameters, we
fixed the surface density exponent $\alpha$ to a value of $-1$ and the grain
size distribution exponent at the canonical value of $-3.5$ (Dohnanyi 1969).
The best fit to the IRS spectrum and \emph{Herschel} detections is found via a Monte Carlo
Markov Chain (\texttt{emcee} package; Foreman-Mackey et al. 2012) on the two
free parameters ($r_{\mathrm{in}}$ and $r_{\mathrm{out}}$). At each iteration,
the total dust mass $M_{\mathrm{dust}}$ is scaled to best match the observed
flux. 
As the primordial disk models described in the previous section,
our debris disk models should be  considered to be highly preliminary.
Due to the sparse SED sampling and lack of spatially resolved data, important
degeneracies remain. In particular, there is an intrinsic degeneracy between 
$r_{\mathrm{in}}$  and the adopted $a_{\mathrm{min}}$ value.
Best fit parameters are reported in Table\,\ref{table:results}.
Note that disk masses are given in M$_{\oplus}$, in contrast to units
of M$_{\mathrm{JUP}}$ used in Table\,\ref{t:pri-fit}.

\section{Discussion}

\subsection{Individual Sources}\label{individual}

\begin{figure}
\includegraphics[width=15cm, trim = 0mm 0mm 0mm 0mm, clip]{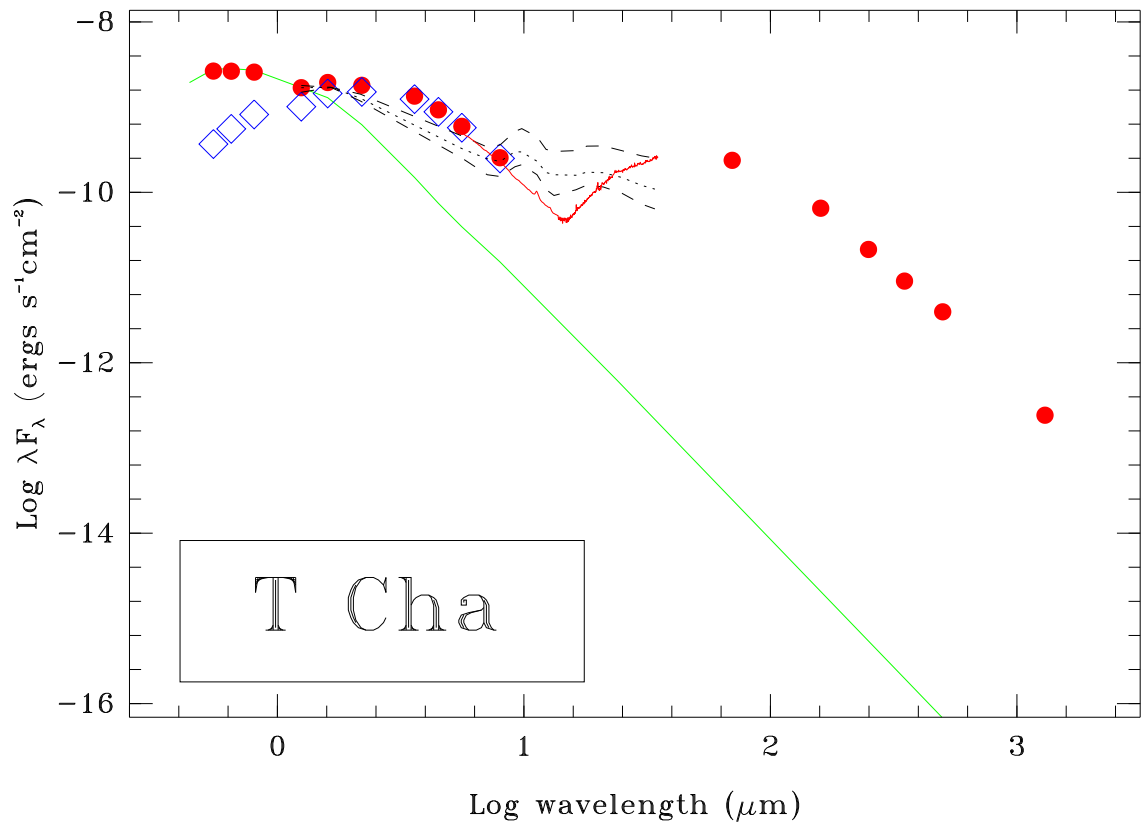}
\caption{The SEDs of  TCha. The symbols are  as in
Figure~\ref{f:border-sed}. The extinction has been estimated from the  R-I
color excess (A$_V$=  4.76$\times$E[R-I]).
The ``dip" in the IRS spectra indicates the presence of a large optically thin
gap separating
optically thick inner and outer disk components. Detailed models have already
been presented
in Cieza et al. (2011).
}
\label{f:tcha-sed}
\end{figure}

\begin{figure}[h]
\begin{center}
\hspace*{-0.cm}\includegraphics[angle=0,width=1.\columnwidth,origin=bl, trim =0mm 0mm 0mm 0mm, clip]{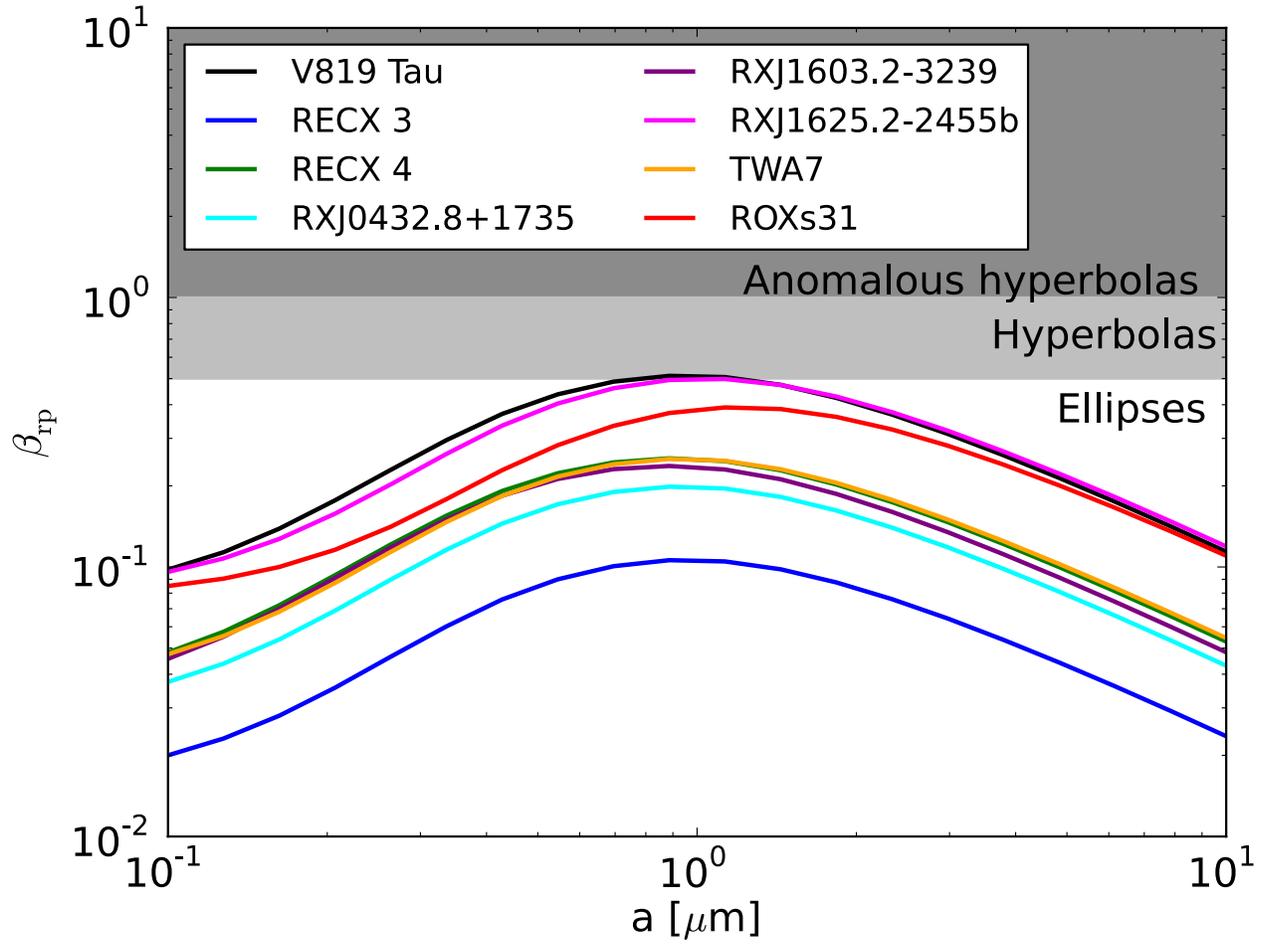}
\caption{\label{fig:beta}Radiation pressure over gravitational forces ratio
$\beta_{\mathrm{rp}}$ as a function of grain size, for the debris disks
sample. As both forces decrease as r$^{-2}$,  $\beta_{\mathrm{rp}}$   is independent of distance
from the star.}
\end{center}
\end{figure}

In this section, we discuss the properties of each of  our targets with IR
excesses.
We divide the sample into different categories,  based on the accretion
signatures and
SEDs, corresponding to the five groups shown in Figures~\ref{f:border-sed}
to~\ref{f:tcha-sed}.

\subsubsection{Border line CTTS/WTTS disks}\label{model-borderline}

The Lupus targets  Sz 68 and Sz 96 have many properties indistinguishable from
typical CTTSs. Their SEDs (see Figure~\ref{f:border-sed}) are
close to the CTTS  median, show no evidence for significant
inner holes, and can be reproduced with relatively standard
disk parameters (see Table~\ref{t:pri-fit}).
Both objects require some amount of dust settling in order to match
their silicate features (i.e., the surface of their disks must be dominated by
small grains,  a few microns in size).
For the Sz 68 disk, we adopt an outer radius of only 15 AU because
it is a close binary system with a \emph{projected} separation of
only 0.13$\arcsec$ (20 AU).
Sz 68 and Sz 96 seem to be borderline objects  moving back and forth across
the CTTS/WTTS boundary.  Hughes et al. (1994)  report a very
small H$\alpha$ EW for Sz 68 (2.8 \AA), while we find
a large H$\alpha$ velocity width (410 km/s).
Conversely,  Hughes et al. (1994) report  a  H$\alpha$ EW of 11 \AA~for Sz 96,
while accretion is barely detectable in our high-resolution spectra
(H$\alpha$ $\Delta$V $\sim$270 km/s, see Figure~\ref{f:halpha}).

\subsubsection{Evolved primordial disks}\label{model-evolved}

Four of our targets,  RXJ1628.2-2405, ROXs 42C, ROXs 47A, and ROXR1 13 seem to
still be accreting circumstellar material, but show very significant decrements
of near-IR excess with respect to typical
CTTSs (see Figure~\ref{f:primordial-sed}). We therefore refer to these objects
as
``evolved"  primordial disks.
We modeled RXJ1628.2-2405 as an \emph{extremely}  flat disk, extending from 0.8
to 100 AU, which has an scale height of 0.17 AU at a radius of  50 AU.
However, we note that this extreme geometry could be an artifact of fixing the outer radius 
to 100 AU and a smaller disk could accommodate a  larger scale height. 
We were unable to reproduce the SEDs of  ROXs 47A and ROXs 42C using a
continuous disk model
because their SEDs  are flat or rising at  $\sim$15 $\mu$m, between the 10
and 20 $\mu$m silicate
features.  We find that  models that fit  well the SED shortward of 15 $\mu$m
significantly
underestimate the observed fluxes between $\sim$20 and 160 $\mu$m.
On the other end, models that best match fluxes longward of 20\,$\mu$m do not match the observed near- to mid-IR fluxes.
However, a  good match to their entire SEDs can be obtained by introducing a
discontinuity
in the disks at  $\sim$1.5-3 AU.
The inner disks are very tenuous and contain $\lesssim$ 0.05 Moon masses of
dust.
Therefore, they could also be regarded as cavities  filled with a small amount
of
dust grains.
These models are  clearly not unique, and similarly good fits can
be obtained with smaller inner disks and gaps separating the inner
and outer disks.
In the case of ROXs 42C, we set the radius  of the outer disk to 30 AU because
ROXs 42C
has a close companion with a \emph{projected} separation of
only 0.28$\arcsec$ (35 AU).
In our models, the disk masses of  RXJ1628.2-2405, ROXs 47A, and ROXs 42C are
$\lesssim$ 0.1-0.01 M$_{\mathrm{JUP}}$, even assuming a gas to dust mass ratio of 100.
However,  even in the optically thin limit,  the far-IR emission is highly dependent on both the 
temperature and mass of the disk. This renders the mass estimates less reliable than
those obtained from (sub)millimeter data.   

Both the IRS spectra and the \emph{Herschel} photometry of ROXR1~13 (also known as DoAr~21)
are very contaminated
by extended emission.  Therefore, the modeling of  the observed IR excess would
not be
very meaningful and we do not present a model.  Jensen et al. (2009) obtained
mid-IR sub-arcsecond
images of the system and concluded that there is little material within
$\sim$100 AU from the star.
While a disk with a very large cavity can not be ruled out, they suggest that
the observed
IR excess is best explained by a small photodissociation region surrounding the
star.
Resolved imaging with the Atacama Large Millimeter/submillimeter Array could
clarify the situation.

\subsubsection{Warm debris disks}\label{model-warm}

The targets  V819 Tau, RXJ0432.8+1735,   REXC~3, REXC~4, TWA~7, and RXJ1603.2-3239
have optically thin disks and show no evidence for accretion.  They are thus
consistent with
gas poor debris disks. Since their IR excesses become evident in the mid-IR, we
classify them
as ``warm" debris disks.
Their SEDs are very well reproduced by our simple models (see
Figure~\ref{f:warm-sed}).
They have inner radii ranging from  $\sim$1 to 13 AU  and outer
radii in the 13 to 420 AU range.
They contain 10$^{-3}$ to 10$^{-1}$ M$_{\oplus}$ of dust (see
Table\,\ref{table:results}).
We note that the fraction of low-mass stars with debris disks detectable via
a 24 $\mu$m excess is very low ($<$4~$\%$) in the field (Trilling et al. 2008)
and that
our young and ``warm" debris disks  are particularly interesting because they
are likely to
represent the  initial conditions of the debris disk phenomenon.
Models of planet formation through core accretion predict that the production
of
second-generation dust will first occur at small radii where the formation
timescales for large ($\gtrsim$ 1000 km) planetesimals capable of stirring the
disk
are shorter (Kenyon $\&$ Bromley, 2005).

\subsubsection{Cold Debris disks}\label{model-cold}

RXJ1625.2-2455b and ROXs 31 are the only systems in our sample
without mid-IR excesses that we detected with \emph{Herschel}.
We classify them as  ``cold" debris disks.
Their low incidence (2 detections versus 16 non-detections, 
see \S \ref{non-det-limits})
is somewhat surprising considering that 16$\%$ of
\emph{much} older
solar-type stars show significant 70 $\mu$m excesses  (Trilling et al. 2008)
and might be a result of a delay
 in the production of large planetesimals  (see section~\ref{dust-production}).
Given the number of photometric points tracing the thermal emission of  ``cold" 
dust (1 and 2 data points for RXJ1625.2-2455b and ROXs\,31, respectively), 
the modeling results are to be interpreted very cautiously. 
We modeled the SED of ROXs 31 with a disk extending from $\sim$40 to 360 AU
Similarly, we reproduced the SED of RXJ1625.2-2455b with a disk extending from 600 to  almost  900 AU
(see Figure ~\ref{f:cold-sed} and Table 7).
Additional SED points and/or resolved imaging are much needed to
better constrain the properties of these systems.
In particular, the dimensions of the disk model for RXJ1625.2-2455b are quite extraordinary.
Even though this object was detected as a point source at both 70 and 160 $\mu$m,
its 160 $\mu$m image shows significant extended emission north-west of the source
(see Figure~2), which could be affecting our photometry.  

We also considered the possibility of an accidental alignment with
an extragalactic source being responsible for the far-IR detections 
toward RXJ1625.2-2455b and ROXs 31. However, source counts at 
the relevant flux levels ($\sim$20 mJy at 70 $\mu$m and $\sim$100 mJy 
at 160 $\mu$m) are of the order of 0.015-0.03/arcmin$^{2}$ 
(Dole et al. 2004), rendering a chance alignment  (within a FWHM distance) 
very unlikely (P $\sim$10$^{-5}$ for any given source).

\subsubsection{The T Cha pre-transition disk}

T Cha is a G8 star with a very variable H$\alpha$ profile (Schisano et al.
2009) suggesting
sporadic accretion.  Its SED (see Figure~\ref{f:tcha-sed}) shows a pronounced
dip in the mid-IR that
is characteristic  of the so-called ``pre-transition" disks (Espaillat et al.
2007), which  have
optically thin gaps separating optically thick inner and outer disk components.
A detailed SED model for the T Cha disk, including the DIGIT \emph{Herschel}
data, has already been presented
in Cieza et al. (2011).  In short,  the T Cha SED can be modeled with a very
small inner disk extending from 0.13 to 0.17 AU
and an outer disk with an inner edge located at $\sim$15 AU.  The outer edge of
the outer disk is not well constrained by
the SED as the size and the surface sensitivity profile of the outer disk are
highly degenerate values.
However, T Cha shows a steep spectral slope at \emph{Herschel} wavelengths,
which  favors models of outer disks containing little or
no dust beyond $\sim$40 AU.  We find that the full SED can be modeled equally
well with either an outer disk that is very compact
(only a few AU wide) or a much larger one (300 AU in radius) that  has a very
steep surface density profile due to
the accumulation of material at the outer edge of the gap.

Hu{\'e}lamo et al. (2011) have recently reported the presence of a low-mass
object candidate,
possibly a young planet or a brown dwarf,  within the gap of the T Cha disk.
This putative object is likely to be responsible for the formation of the gap (Dobson-Robinson $\&$ Salyk (2011)
and \emph{could} be related to the peculiar surface density profile of the
outer disk \emph{if} future resolved submillimeter images of the T Cha system
confirm
that material is accumulating at the outer edge of the gap.

\subsubsection{Non-detections}\label{non-det-limits}

\begin{figure}
\includegraphics[width=8cm, trim = 0mm 0mm 0mm 0mm, clip]{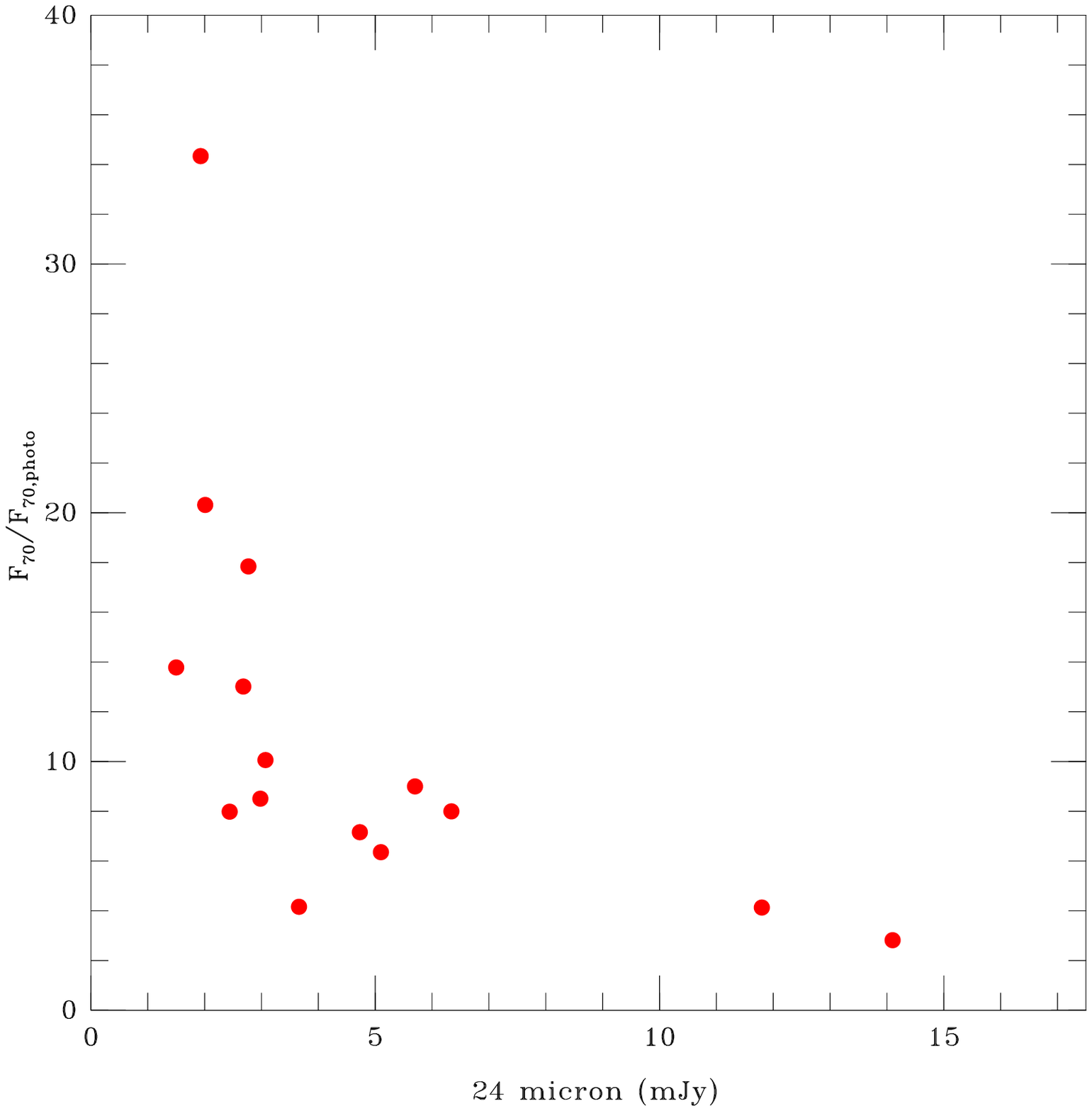}
\includegraphics[width=8cm, trim = 0mm 0mm 0mm 0mm, clip]{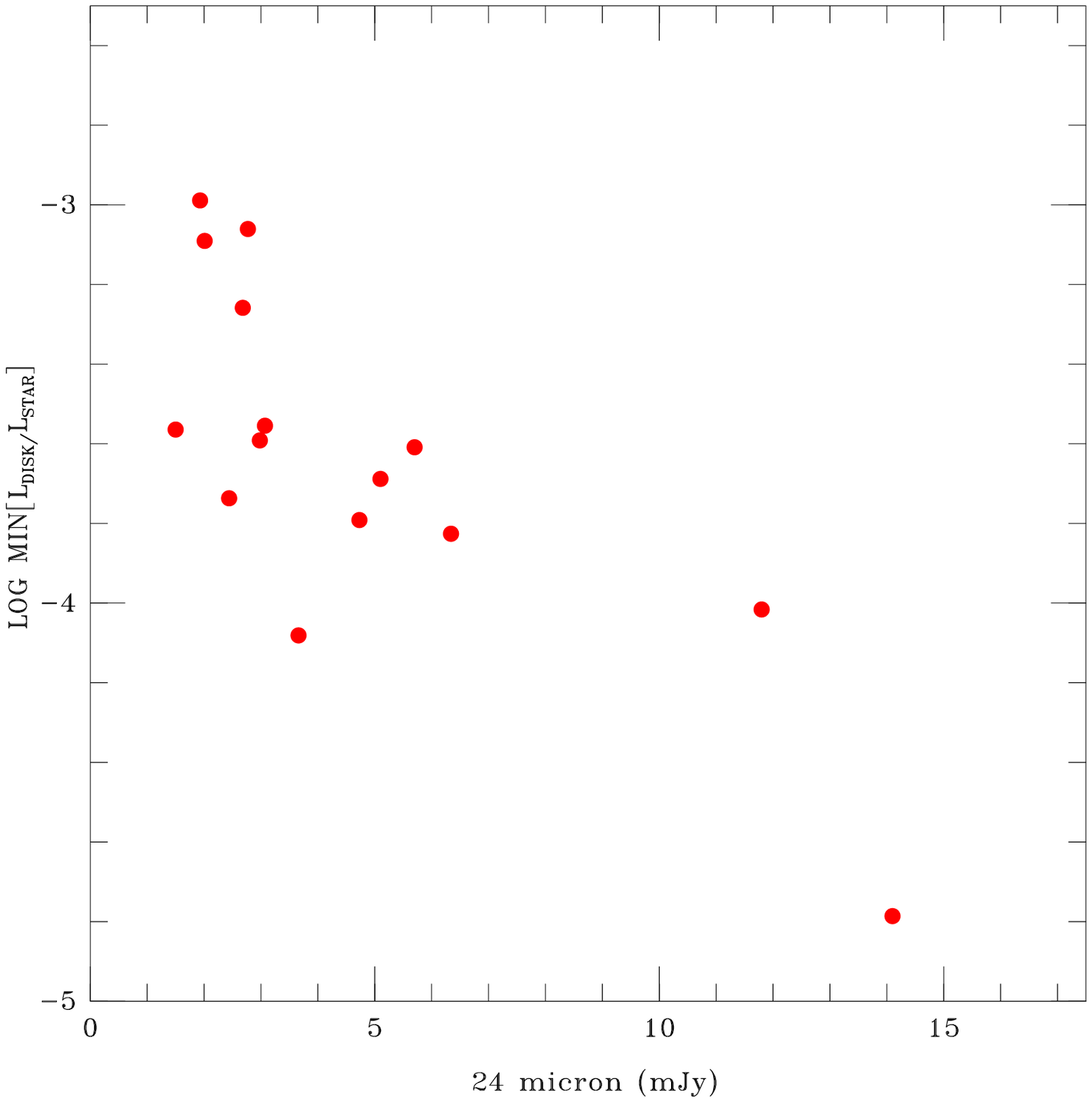}
\caption{\textbf{Left panel:} the ratio of the  3--$\sigma$ 70 $\mu$m upper
limit to the expected
photospheric  flux  at 70 $\mu$m as a function of  photospheric 24 $\mu$m flux
for
the diskless targets.
For most of the targets, we are sensitive to  disks with 70 $\mu$m fluxes that
are
$\sim$5-15 times higher than the stellar photospheres.
\textbf{Right panel:} the 70 $\mu$m 3-$\sigma$ upper limits translated into
fractional disk luminosity
limits as described in Section~\ref{non-det-limits}. These limits correspond to
$\sim$50 K disks whose emission
peaks at 70 $\mu$m.
}
\label{f:f70}
\end{figure}

One important result from our survey is that objects with photospheric 24
$\mu$m fluxes
but large far-IR and submillimeter excesses are relatively rare.  In our sample
of 18 targets without
\emph{Spitzer} excesses, there are only two objects with \emph{Herschel}
detections,
RXJ1625.2-2455b and ROXs 31. This implies that the overall disk fraction of
WTTSs
in nearby molecular clouds and young stellar associations is not much larger
than  the
$\sim$20$\%$ value estimated from \emph{Spitzer}  observations (Cieza et al.
2007, Wahhaj et al. 2011).
The vast majority of WTTSs do in fact seem to be diskless;
however, the  presence of a disk can only be ruled out within certain limits
given by
the depth and/or  precision of the observations.
In the case of our survey, the 70 $\mu$m data are the most constraining.
Therefore,
as a first step to establish the limits of our survey,  we plot the  70 $\mu$m
upper 3-$\sigma$ limits
divided by the expected photospheric value (calculated by extrapolating the
observed mid-IR fluxes)
as a function of the photospheric 24 $\mu$m  flux. As shown in
Figure~\ref{f:f70} (left panel), for most of
our targets, our observations are sensitive  to disks with 70 $\mu$m fluxes
$\sim$5-15 times higher than the
stellar photospheres.

These 3-$\sigma$ upper limits can be translated into  minimum detectable disk
luminosities, $L_{disk}/L_*$,
 by setting the emission peak at 70\,$\mu$m (T$_{disk} = 52.5 K)$.
Following Bryden et al.  (2006), we calculate minimum disk luminosity
as a function of 70\,$\mu$m  upper limit, according to:

\begin{equation}\label{FDL}
\frac{L_{disk}}{L_*}(\mathrm{minimum})
= 10^{-5}\left(\frac{5600 K}{T_{*}}\right)^3
\left(\frac{3\sigma_{70}-F_{*,70}}{F_{*,70}}\right)
\label{eq1}
\end{equation} \\
where  $F_{*,70}$ is the expected  flux of the star  at 70\,$\mu$m.
The minimum $L_{diskt}/L_*$ values that would be detectable
in our  sample are shown in Figure~\ref{f:f70} (right panel).
For most of  our non-detections,  disks fainter than $L_{disk}/L_*$ $\sim$
10$^{-3}$--10$^{-4}$ can
not be ruled out.
Given the strong temperature dependence in Equation~\ref{eq1},  it is clear that
the late types of the stars in our sample (mostly late K and M-type stars), could
be a contributing factor to the high rate of non-detections. 

\begin{figure}[h]
\begin{center}
\hspace*{-0.cm}\includegraphics[angle=0,width=1.\columnwidth,origin=bl, trim =0mm 0mm 0mm 0mm, clip]{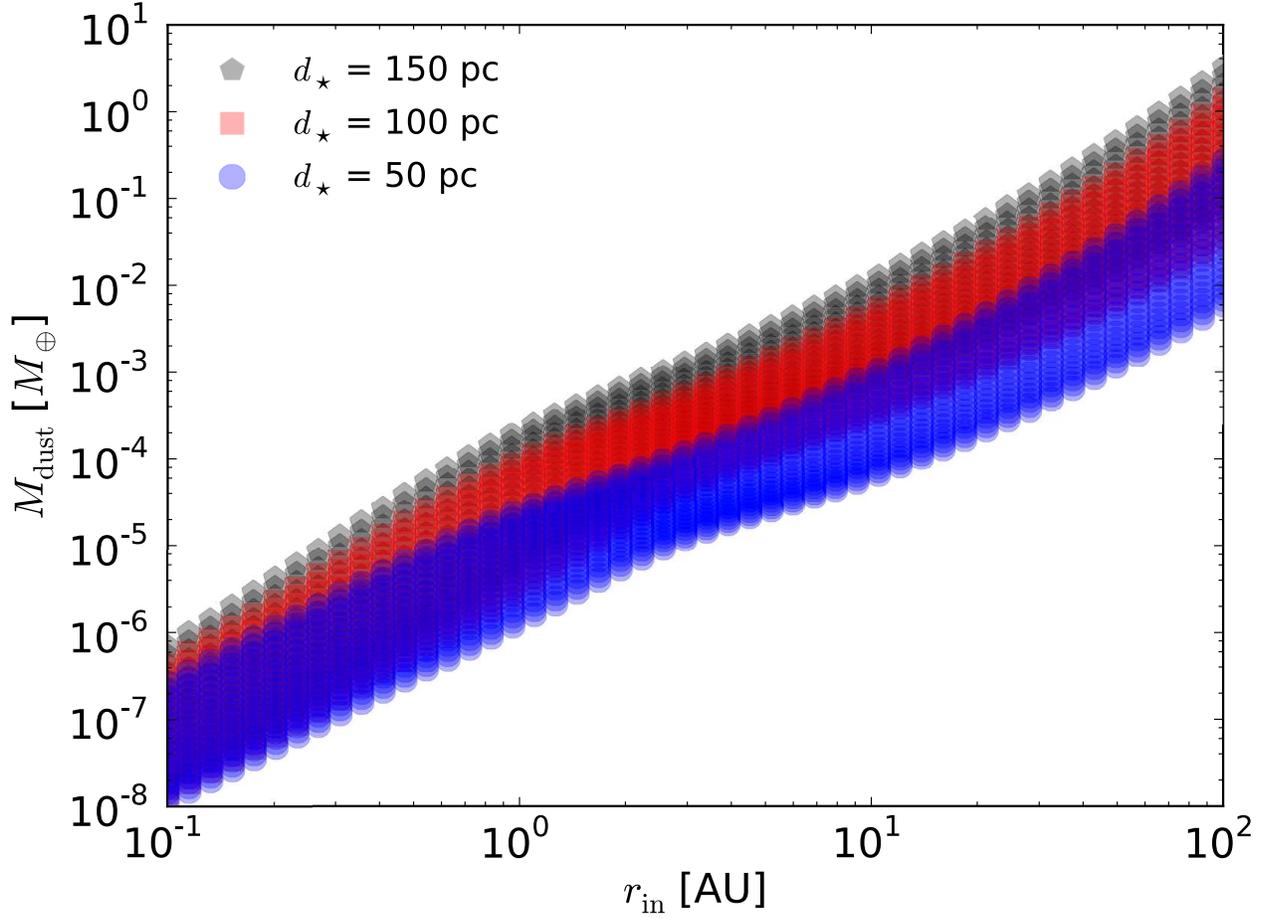}
\caption{\label{fig:masses}Model-derived upper limits on the encompassed dust mass
$M_{\mathrm{dust}}$ (within $r_{\mathrm{in}}$) as a function of distance to that star
that could  still lead to a non-detection with our observations.}
\end{center}
\end{figure}

Converting the \emph{Herschel} upper limits into dust mass constraints
requires some modeling.
Following Cieza et al. (2007; 2008b),   we use the optically thin disk models
discussed in Section~\ref{thin}
to constrain the maximum amount of dust that could remain undetected within the
first $\sim$100 AU of the apparently diskless stars in our WTTS sample.
Figure\,\ref{fig:masses} shows the upper limits on the dust mass that can be
hidden below our non-detections. We adopted the SED of the object RECX\,6
($d_{\star} = 100$\,pc), which is representative of our stellar sample (M0-type
star), and ran a grid of models. We took 50 values for $r_{\mathrm{in}}$,
logarithmically spaced between 0.1 and 100\,AU. For the width $W$ of the disk
($r_{\mathrm out}  = W \times r_{\mathrm in}$) , we took 20 values,
logarithmically spaced between 1.1 and 10. The 70\,$\mu$m flux is the most
sensitive among our observations, and we adopted a 3\,$\sigma$ upper limit of
3\,mJy. For each model of the grid, we scale the dust mass in order to reach
this 3\,$\sigma$ flux level. We then repeated this exercise by scaling up and
down the distance of RECX\,6 to 150 and 50\,pc, respectively, which are the
maximum and minimum distances in our stellar sample (except for Sz 96 at 200
pc). The fluxes were scaled accordingly to the new distances. In
Figure\,\ref{fig:masses},  blue circles, red squares, and black pentagons show
the results of the three grids, for distances of 50, 100, and 150\,pc,
respectively. The spread in $M_{\mathrm{dust}}$ reflects the different values
for the width $W$ of the disk.

For objects at 100 pc, we are sensitive to $\sim$10$^{-5}$, 10$^{-3}$, and 10$^{-1}$
M$_{\oplus}$ of dust within 1, 10, and 100 AU from the central star.
Of course, these limits only apply for dust grains with sizes between
$a_{\mathrm{min}} =  0.1$\,$\mu$m  and $a_{\mathrm{max}}=1$\,mm.
Our models provide no constraints on the mass of solids that could be
present in the form of larger bodies (i.e., planetesimals or planets) because
their contribution to the opacity of the disk is negligible.

\subsection{The broader context of disk evolution and
dissipation}\label{evolution}

In Section~\ref{individual}, we have grouped our detected disks into several
categories
based on their SEDs and accretion properties. In this section, we discuss how
the objects from each category might fit into  the broader context of disk evolution and
dissipation.
As discussed in the introduction, protoplanetary disks evolve through a variety
of physical processes,
including viscous accretion, grain growth and dust settling, photoevaporation,
and the dynamical interaction with stellar and substellar companions.
The relative importance of all these processes is still not well understood,  
but the observational evidence and theoretical work seem to support
the following emerging picture.

\subsubsection{Primordial disk evolution}

Early in their evolution, protoplanetary disks lose  mass through
accretion onto the star (Hartman et al. 1998), outflows, and photevaporation
by UV photons and X-rays (Gorti et al. 2009).
During this ``mass depletion"  stage, objects are classified as CTTSs based
on their clear accretion signatures, and they maintain more or less constant IR
SEDs
as the entire disk remains optically thick to IR radiation. The main
observational sign of evolution at this point is a 
rapid  reduction in the
(sub)millimeter fluxes with age as the mass of the disk is depleted and/or the
opacity
of the dust decreases due to grain growth (Rodmann et al. 2006; Lee et al.  2011)
Depending on the initial conditions, the CTTS phase can last for up
to 5 to 10 Myrs, but it can be much shorter ($<$ 1 Myr) in some cases. 
For example, disks in close binary systems disappear more rapidly
(Cieza et al. 2009; Kraus et al. 2011), and
different initial core properties and evolution during the embedded phase
can affect the disk evolution (Dunham $\&$ Vorobyov 2012). 
Objects Sz 68 and Sz 96 are likely to represent late stages of the ``mass
depletion" stage.
They still retain full disks with SEDs  very close to the CTTS median,
but these disks are not very massive ($\lesssim$ 1-4 M$_{\mathrm{JUP}}$).

During the CTTS phase, dust settling steepens the slope of  the mid-IR 
(Dullemond $\&$ Dominik, 2004) and grain growth may generate inner opacity holes  (Dullemond $\&$ Dominik, 2005).
Also, as the disk mass and accretion rate decrease,
the accretion rate drops below the photoevaporation rate
and the inner disk drains on a viscous timescale.
The combination of grain-growth, dust settling, and/or photoevaporation can
explain
the properties of  the disks around  RXJ1628.2-2405, ROXs 47A, ROXs 42C, and
(perhaps) ROXR1 13.
All these evolved primordial disks have very low masses ($\lesssim$ 0.1
M$_{\mathrm{JUP}}$) and strong
evidence for inner holes and/or dust settling.

\subsubsection{The primordial to debris disk transition}

The formation of an inner hole through photoevaporation  stops accretion 
and  marks the rapid transition between the CTTS and the WTTS stage.
At this point,  the gaseous disk quickly photoevaporates from the inside out, leaving 
behind solid objects, which  may collide with each other producing the
second-generation dust seen in debris disks. 
The eight objects shown in Figures~\ref{f:warm-sed} and~\ref{f:cold-sed}
seem to be past the stage when the inner disk has drained
completely and  photoevaporation prevents any further accretion onto the star.
Whether their \emph{outer} disks still contain any remaining gas or they are
truly gas-poor debris disks is currently unknown.

\begin{figure}[h]
\begin{center}
\hspace*{-0.cm}\includegraphics[angle=0,width=1.\columnwidth,origin=bl, trim =0mm 0mm 0mm 0mm, clip]{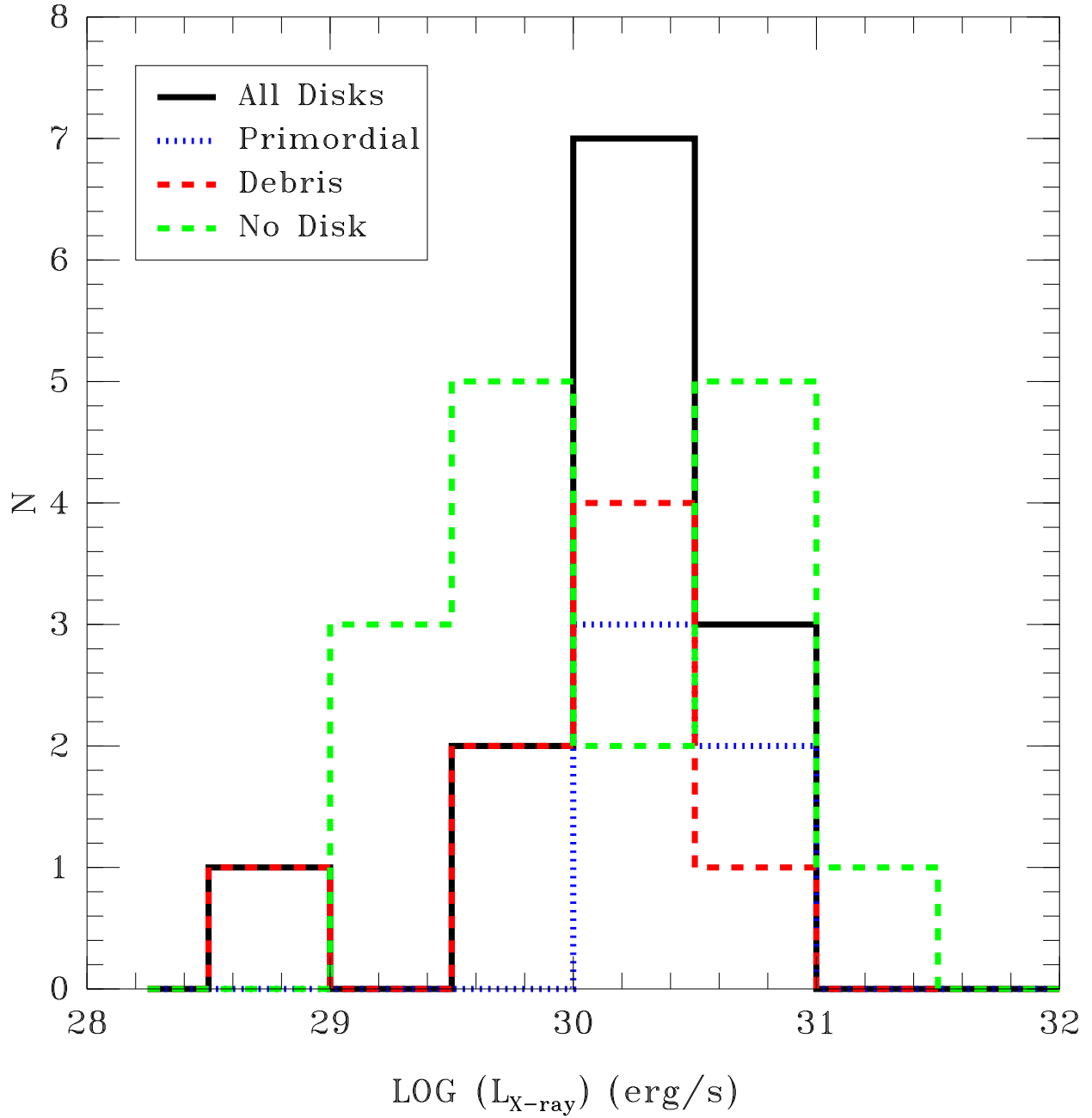}
\caption{\label{fig:xray} The distribution of X-ray luminosities in our sample
of WTTS,
divided into different categories. In our small sample, no statistically
significant evidence is seen between any group.
In particular, the X-ray luminosities of  the targets without detected disks do not seem 
to be systematically higher than those of the rest of the sample.
}
\end{center}
\end{figure}

While it is clear that photoevaporation 
plays a central role in the final dispersal of the primordial disk and the transition
to
the debris disk stage,  the dominant mechanism is not well established.
EUV  photoevaporation models (Alexander et al. 2006) predict low
evaporation rates
 of the order of 10$^{-10}$ M$_{\odot}$yr$^{-1}$ and should  only become
important once the accretion rate has fallen below this level.
More recent  FUV+X-ray photoevaporation models
(Owen et al. 2011) predict much higher evaporation rates,
which could reach levels of the order of 10$^{-8}$-10$^{-7}$
M$_{\odot}$yr$^{-1}$
and scale linearly with X-ray luminosity.  Such large evaporation rates may be
higher than the
accretion rates of typical CTTSs and could drive the dissipation of the disk
at early stages.
Under such circumstances, one may expect to find an anti-correlation 
between X-ray luminosity and
the presence of a primordial disk since the circumstellar gas should evaporate
faster around the most
X-ray luminous stars.
Figure~\ref{fig:xray} shows that the distribution of X-ray luminosities (taken
from Liebhart et al.,  in preparation)
of our sample of WTTS divided into different categories: stars with disks,
stars with primordial disks,
stars with debris disks, and stars without a disk.
No statistically significant difference is seen between \emph{any} of the
groups.
This suggests that X-ray luminosity is not a dominant factor in disk evolution,
possibly  because the X-ray photoevaporation rates are lower than
predicted by
recent models or because other factors, such as the variability and evolution
of the X-ray luminosity, and the mass, multiplicity, and  age of the stars,
are masking the signal in our small sample.

\subsubsection{Alternative SED evolutionary paths}

T Cha does not fit  the above picture in which primordial disks
remain optically thick in the IR at all radii until the entire disk suddenly
dissipates
from the inside out through photoevaporation. Instead,
the gap in the T Cha disk is best explained by the dynamical interaction of one
or more
substellar objects embedded within the disk.
The potential direct detection of such an object has been reported by Hu{\'e}lamo et al. (2011),
but awaits confirmation. 
Recent hydrodynamical simulations of giant planets embedded
in primordial disks by Dodson-Robinson \& Salyk (2011)
and Zhu et al. (2011) show that multiple giant planets are
needed to produce gaps as wide as the  one inferred for T Cha.
Since pre-transition disks with obvious gaps like T Cha are very
rare, they could  represent special cases, where multiple massive planets may
be present, rather than young solar system analogs with a single Jupiter-mass
planet.

\subsubsection{Non-detections, the formation of planetesimals, and dust
production}\label{dust-production}

The large number of debris disks known to surround stars much older than the T Tauri stars studied here suggests that 
planetesimal formation is efficient and common, even though the formation mechanism  for such planetesimals 
remains uncertain  (see Chiang $\&$ Youdin, 2010 for a recent review). As the abundance of extrasolar giant planets known to date 
(Howard et al. 2012)  further suggests that planetesimals must form within a Myr if core accretion is to operate within 
the disk gas dissipation timescale, the majority of the disks in our sample should have formed planetesimals. 
However, half of our targets show no excess at any wavelength. If their disks have formed planetesimals, 
what is limiting the debris detectability in these systems?

Models that attempt to fit the \emph{Spitzer} excesses of  nearby debris disks
around
older solar type stars can be extrapolated back to the 1-10 Myr
ages of our sample for comparison with our results.
For instance, Kains, Wyatt, $\&$ Greaves (2011)
assume a planetesimal disk in collisional equilibrium. Their model
predicts  L$_{disk}$/L$_{\star}$ = 10$^{-2}$ to 10$^{-6}$  at an
age of 1 Myr for initial solid masses $\sim$3-300 M$_{\oplus}$.
Since the lower end of the predicted L$_{disk}$/L$_{\star}$ distribution is
below our sensitivity limits calculated in Section~\ref{non-det-limits},
we conclude that our \emph{Herschel} observations are likely to miss
many of the disks with lower initial solid masses and/or larger radii. 
Another possibility for the relatively low detection rate of debris disks  is
the existence of a  quiescent phase between the dissipation
of the primordial disk an the onset of the debris disk phenomenon.
Models by  Kenyon $\&$ Bromley (2008)
consider how planetesimals grow and obtain their eccentricities.
In their models of planetesimals at 30--150 AU,  eccentricities start out
very small and reach 0.05 only at  $\sim$100 Myr. As a result, significant
dust production is delayed until $\sim$5-10 Myr and peaks around 10-50 Myr.
We should also consider the possibility that the planetesimal disks in the
 systems without \emph{Herschel} excesses have been removed as
a consequence of giant planet formation. Raymond et al. (2012)
 explore the dynamical evolution of and
 dust production in disks that have formed multiple giant planets,
planetesimals, and planetary embryos.
They find that systems in which the giant
planets undergo a strong dynamical instability can clear planetesimals
from both the inner terrestrial planet region and the Kuiper Belt
region of the disk, limiting future debris production.  These
systems end up with giant planets on eccentric orbits and are
unlikely to have terrestrial planets.
The  data currently available does not distinguish between the three
potential explanations for the non-detections discussed above.

\subsubsection{Stellar mass, age, and  multiplicity}

Our small sample does not show statistically significant trends in the disk
fractions or properties as a function of  stellar mass, multiplicity,
or membership in regions  with different mean ages.
The one exception is perhaps the low incidence of primordial disks among the
oldest
targets in our sample. Nine of  our targets are in regions with estimated ages
$\gtrsim$ 5 Myr
(TWA,  $\eta$ Cha and $\epsilon$ Cha; Torres et al. 2008). Only one (11$\%$) of
them, T Cha, harbors an accreting
primordial disk.  In contrast, we find 6 gas-rich primordial disks among the 22
targets (i.e., 27$\%$)
from the younger 1--3 Myr  regions (Ophiuchus, Lupus, and Taurus; Comeron 2008,
Wilking et al. 2008, Kenyon et al. 2008).
Nevertheless, these young regions also contain many debris disks  and diskless
stars.
This is not surprising considering that disk evolution is a function of many
variables besides age, including
stellar mass, multiplicity, initial  conditions, and environment (Williams $\&$
Cieza, 2011).
Disentangling the effects from each variable requires very large samples
and/or  further parametric constraints. 

\subsection{Outstanding questions}

The IR SEDs of the WTTSs in our sample trace the late evolution of primordial
disks and the transition to a debris disk  or an apparently diskless stage.
Our modeling work places constraints on the  dust content  and disk geometry of
our targets.
However, important disk evolution questions still remain unanswered.
In particular, the gas content of our objects is very poorly established.
The high-resolution data of the H$\alpha$ line can unambiguously
identify the presence of accretion and thus a gas-rich disk. Nevertheless,
the lack of accretion signatures does not necessarily imply
a gas-poor disk. Accretion rates are highly variable and
can fall below detectable levels.  Also, processes such as photoevaporation
and the dynamical interaction of embedded  objects can prevent accretion
from a gas-rich disk.
Furthermore, detecting accretion  provides little to no information on the
\emph{total amount} of circumstellar gas present in the disk.

Establishing the evolution of the gas (and thereby of the gas to dust mass
ratio) in
protoplanetary disks is currently one of the biggest  challenges in the field.
The gas content is  critical for the formation of not only giant planets, but
also
terrestrial planets because it controls the dynamics of dust grains, the
building blocks
of rocky objects.
While most  disk models assume a gas to dust mass ratio of 100 (the canonical
value for
the interstellar medium), it is clear that this is \emph{not} a particularly
safe assumption.
On the one hand, gas photevaporation will tend to decrease the gas to dust mass
ratio with time.
On the other hand, this ratio should increase as grains grow into planetesimals
and planets.
At the end of the primordial disk dissipation process, many (or most)
protoplanetary disks should evolve
into dusty gas-poor debris disks,  even if these tenuous disks are not bright
enough to be detectable by current surveys
(e.g.,  like the debris disk in our own solar system).
However, the evolution of the gas  to dust mass ratio could be complex  rather
than follow a monotonically
decreasing function.  Disks with gas to dust mass ratios significantly greater
than 100 can thus not be ruled out
(Pinte et al. 2010).
To complicate things, the evolution of the gas to dust mass ratio is expected
to be dependent on variables such
as stellar mass and the planet formation history of any given system.

Obtaining reliable estimates for the gas content of protoplanetary disks will
require observations of
several gas tracers from different regions of the disk and detailed models of
the disk structure.
Spatially resolved observations of the continuum and multiple molecular species
and isotopologues
 with ALMA
 will help tremendously in this area.
Reliable estimates of the gas to dust mass ratios and disk structures of all
our WTTS disks would
result in a much clearer view of the disk evolution picture discussed in
Section~\ref{evolution}.
In particular, they would help establish whether the optically thin disks in our sample 
are the remnants of primordial disks are true gas-poor debris disks where the opacity 
is dominated by second-generation dust produced by the collation of planetesimals.

\section{Summary and conclusions}

In this study, part of the  DIGIT  \emph{Herschel}  Key Program,
we presented  far-IR and submillimeter photometry  of 31 WTTSs in order to
investigate
the properties and evolutionary status of their disks.
We constructed preliminary disk  models for all the objects with IR excesses.
Our main conclusions can be summarized as follows:

\noindent 1) WTTSs are a very diverse population in terms of SEDs.
In our sample,  6$\%$ (2/31)  of the targets exhibit IR excesses
indistinguishable from  those of CTTSs
and are likely to be borderline objects moving back and forth across the
observational boundary between CTTSs and WTTSs.
13$\%$ (4/31) of our WTTSs have SEDs consistent with evolved primordial disks,
while 26$\%$  (8/31) of them  seem to be in a debris disk stage.
One target, T Cha,  is a ``pre-transition" disk  with gap in the disk and
intriguing outer
disk properties.
Half (16/31) of our targets lack  detectable levels of near-, mid-, or far-IR
excesses.

\noindent 2) Objects without excess 24 $\mu$m emission over that from 
the photosphere, but detectable
\emph{Herschel} excesses are rare,  $\sim$11$\%$ (2/18) in our sample of
objects without \emph{Spitzer} excesses.
This implies that the overall disk fraction of  WTTSs is not much higher than
$\sim$20$\%$, the value obtained by  previous  \emph{Spitzer} surveys at
$\lambda \lesssim$ 24 $\mu$m.
However,  since our \emph{Herschel} observations are not sensitive to
the stellar photospheres of our targets, this conclusion is only valid for
disks
where $F_{70}$/$F_{70,\star}$ $\gtrsim$ 5--15 and
 $L_{disk}$/$L_{\star}$ $\gtrsim$ 10$^{-3}$--10$^{-4}$.

\noindent 3)
Using the velocity dispersion of the H$\alpha$ line, which is
a more sensitive accretion indicator than H$\alpha$  EWs, we find that
all the 7 targets with optically thick disks are in fact weakly accreting
objects (i.e.,
primordial disks).  In contrast,  none of the 8 optically thin disks show
evidence for accretion, and are most likely to be debris disks.
These results support the idea that the transition from the primordial to
debris disk phase happens very quickly  through photoevaporation.
This transition occurs when most of the disk mass has been depleted and the
accretion rate
falls below the photovaporation rates, at which point the gaseous disk
dissipates  from the inside out, leaving behind large grains, planetesimals
and/or planets.

\noindent 4)
T Cha has a relatively massive disk  with a $\sim$15 AU wide  gap
separating optically thick  inner and outer disk components.  This object
is an exception to the  emerging picture in which most protoplanetary disks
remain optically thick in the IR until the entire disk suddenly dissipates
from the inside out through photoevaporation. Instead, the
properties of T Cha are best explained by the formation of
multiple planets  massive enough to open wide, overlapping
gaps in the disk.

\noindent 5)  Preliminary SED modeling of  the optically thick  disks  in our sample
shows  significant evolution in disk properties with respect to typical
CTTS disks, including reduced disk masses, evidence for extreme
dust settling, and the presence of inner holes and gaps.

\noindent 6) Preliminary SED modeling of the optically thin objects suggests that
young debris disks around WTTSs tend to be warmer than older analogs
seen in the field. This is consistent with ``delayed stirring" models
of the production of second-generation dust. However, two important
caveats remain:  WTTSs have not yet been observed in the far-IR to the same
mass sensitivity limits  as nearby MS stars and the debris disk /gas-poor
status of young optically thin disks still needs be confirmed.
\acknowledgments
Support for this work, part of the DIGIT \emph{Herschel} Open Time Key
Program, was provided by NASA through an award issued by JPL/Caltech.
LAC acknowledges support from  NASA through
the \emph{Sagan} Fellowship Program.
CP acknowledges funding from the 
European Commission's 7$^\mathrm{th}$ Framework Program 
(contract PERG06-GA-2009-256513) and from 
Agence Nationale pour la Recherche (ANR) of France under contract
ANR-2010-JCJC-0504-01. 

\emph{Facilities}:  {\it{Herschel}} (PACS, SPIRE), and
CFHT (Espadons).

\begin{deluxetable}{lrrlrclcccccc}
\tablewidth{0pt}
\tablecaption{\emph{Sample Properties}}
\tabletypesize{\tiny}
\tablehead{
\colhead{ID}&\colhead{RA}&\colhead{DEC}&\colhead{Region}&\colhead{Dist}&\colhead{Ref\tablenotemark{1}}&\colhead{SpT}&\colhead{H$\alpha$ EW}&\colhead{Ref\tablenotemark{2}}&\colhead{$\Delta$V H$\alpha$ }&\colhead{Ref\tablenotemark{3}}&\colhead{Bin. sep.\tablenotemark{4}}&\colhead{Ref\tablenotemark{5}}\\
\colhead{}&\colhead{J2000.0}&\colhead{(J2000.0)}&\colhead{}&\colhead{(pc)}&\colhead{}&\colhead{}&\colhead{(\AA)}&\colhead{}&\colhead{(km/s)}&\colhead{}&\colhead{(\arcsec)}  
 }
\startdata
V1096 Tau &   04h13m27.21s & +28d16m24.7s                    &    Tau  &    140   &    1  &  M0    &  3.0    &  H88  &\nodata&\nodata& 0.015   &  K11\\
Hubble 4  &  04h18m47.02s & +28d20m08.4s                   &    Tau  &    140   &      1  &  K7    &  3.0    &  H88  &   188 & N12   & 0.028   &  K11\\
V819 Tau  &  04h19m26.27s & +28d26m14.2s                   &    Tau  &    140   &      1  &  K7    &  1.7    &  H88  &   180 & C12   & noVB/SB &  L93,N12 \\
HD283572  &  04h21m58.86s & +28d18m06.5s                   &    Tau  &    140   &      1  &  G2    & -0.63   &  W10  &  -187 & W10   & noVB/SB &  L93,N12 \\
L1551-51  &  04h32m09.28s & +17d57m23.3s                   &    Tau  &    140   &      1  &  K7    &  0.56   &  W10  &   114 & W10   & noVB/SB &  K11,N12 \\
V827 Tau  &  04h32m14.56s & +18d20m15.0s                   &    Tau  &    140   &      1  &  K7    &  1.8    &  H88  &   168 & N12   & 0.093   &  N12 \\
RXJ0432.8+1735  & 04h32m53.24s & +17d35m33.7s              &    Tau  &    140   &      1  &  M2    &  1.9    &  M99  &   138 & W10   & noSB    &  N12 \\
V830 Tau  &   04h33m10.04s & +24d33m43.3s                  &    Tau  &    140   &      1  &  K7    &  3.0    &  H88  &   121 &  N12  & noVB/SB &  K11,N12 \\
RECX 3  &   08h41m37.03s & -79d03m30.4s                    &$\eta$ Cha&   100   &      2  &  M3    &  2.2    &  S09  &   116 & J06   & noVB    &  K02 \\
RECX 4  &   08h42m23.77s & -79d04m03.0s                    &$\eta$ Cha&   100   &      2  &  M1.5  &  2.3    &  S09  &   147 & J06   & noVB    &  K02 \\
RECX 6  &   08h42m38.77s & -78d54m42.7s                    &$\eta$ Cha&   100   &      2  &  M3    &  3.6    &  S09  &   145 & J06   & noVB    &  K02 \\
RECX 10  &   08h44m31.90s & -78d46m31.1s                   &$\eta$ Cha&   100   &      2  &  M0.5  &  1.0    &  S09  &   103 & J06   & noVB    &  K02 \\
TWA 6  &   10h18m28.70s & -31d50m02.9s                     &     TWA &     50   &      2  &  M0    &  3.8    &  T06  &\nodata&\nodata& noVB    &  E12 \\
TWA 7  &   10h42m30.11s & -33d40m16.2s                     &      TWA&     50   &      2  &  M2    &  4.9    &  T06  &   109 &  J06  & noVB    &  E12\\
T Cha  &   11h57m13.53s & -79d21m31.5s                     &$\epsilon$ Cha&100  &      2  &  G8    &  0.3-30 &  SH9  &   450 &  SH9  & PP      &  H11 \\     
TWA 10  &   12h35m04.25s & -41d36m38.6s                    &     TWA &     50   &      2  &  M2    &  4.8    &  T06  &   199 &  J06  & noBV    &  E12\\
TWA 17  &   13h20m45.39s & -46d11m37.7s                    &     TWA &     50   &      2  &  K5    &  2.5    &  R03  &\nodata&\nodata& SB      &  J06\\
Sz 67  &   15h40m38.27s & -34d21m36.4s                     &  Lup  I &    150   &      3  &  M4    &  5.9    &  H94  &\nodata&\nodata&\nodata  & \nodata\\
Sz 68  &   15h45m12.87s & -34d17m30.6s                     &  Lup  I &    150   &      3  &  K2    &  2.8    &  H94  &  410  &  TW   & 0.126   & C06 \\
RXJ1603.2-3239  &  16h03m11.82s & -32d39m20.2s             &  Lup  I &    150   &      3  &  K7    &  1.1    &  K97  &  132  &  W10  &\nodata  & \nodata\\
Sz 96  &   16h08m12.64s & -39d08m33.3s                     & Lup III &    200   &      3  &  M1.5  &   11    &  H94  &  270  &  TW   & noSB    & M03 \\
RXJ1622.6-2345  &   16h22m37.55s & -23d45m50.4s            &    Oph  &    130   &      4  &  M2.5  &  3.6    &  M98  &  128  & W10   & noSB    & P07 \\
RXJ1625.2-2455b  &   16h25m14.69s & -24d56m07.1s           &    Oph  &    130   &      4  &  M0    &  2.9    &  M98  &  206  & W10   & noSB    & P07 \\
ROXs 3  &   16h25m49.64s & -24d51m31.9s                    &    Oph  &    130   &      4  &  M0    &  2.8    &  B92  &\nodata&\nodata& noVB    & R05 \\
ROXR1 13  &   16h26m03.03s & -24d23m36.4s                  &    Oph  &    130   &      4  &  K0    &  0.7    &  M98  & 400   &  J09  & 0.005   & L08 \\
RXJ1627.2-2410  &   16h27m11.90s & -24d10m31.1s            &    Oph  &    130   &      4  &  M0    &  4.5    &  M98  &\nodata&\nodata&\nodata &\nodata  \\
ROXs 31  &   16h27m52.09s & -24d40m50.2s                   &    Oph  &    130   &      4  &  K7    &  3.3    &  B92  &\nodata&\nodata& 0.480 & R05 \\
RXJ1628.2-2405  &   16h28m16.74s & -24d05m14.2s            &    Oph  &    130   &      4  &  K5    &  3.3    &  M98  & 470   &  TW   & noVB  & R05 \\
ROXs 39  &   16h30m35.63s & -24d34m18.6s                   &    Oph  &    130   &      4  &  K5    &  6.0    &  B92  &\nodata&\nodata& noVB  & R05 \\
ROXs 42C  &   16h31m15.79s & -24d34m01.9s                  &    Oph  &    130   &      4  &  K6    &  1.6    &  B92  &  390  &  TW   & 0.277 & R05 \\
ROXs 47A  &   16h32m11.80s & -24d40m21.8s                  &    Oph  &    130   &      4  &  K5    &  9.2    &  B92  &  430  &  TW   & 0.784 & R05  
\enddata
\tablenotetext{1}{
References for distances are: 
1 = Kenyon, Gomes \& Whitney, 2008;
2 = Torres et al. 2008; 
3 = Comeron 2008;
4 = Wilking, Gagne, \& Allen, 2008.
}
\tablenotetext{2}{
References for spectral types and H$\alpha$ EW are:
B92 = Bouvier $\&$ Appenzeller, 1992;  
H88 = Herbig $\&$ Bell, 1988;
H94 = Hughes et al. 1994;
K97 = Krautter et al. 1997;
M99 = Martin $\&$ Magazzu, 1999;
M98 = Martin et al. 1998;
R03 = Reid et al. 2003;
S09 = Sicilia-Aguilar et al. 2009;
SH9 = Schisano et al. 2009; 
T06 = Torres et al. 2006;
W10 = Wahhaj etl at. 2010.
}
\tablenotetext{3}{
References for H$\alpha$ velocity widths are: 
C12 =Cieza et al. 2012;
J06 =Jayawardhana et al. 2006;
J09 =Jensen et al. 2009;
N12= Nguyen et al. 2012;
SH9 = Schisano et al. 2009;
TW = this work;
W10= Wahhaj etl at. 2010.
}
\tablenotetext{4}{Binary separation, SB = spectroscopic binary, no VB = no visual binary within survey limits; no SB = no spectroscopic binary within survey limits; PP = possible planet embedded in the disk.}
\tablenotetext{5}{
References for binarity are:
C06 = Correia et al. (2006);
E12 = Evans et al. (2012);
J06 = Jayawardhana et al. (2006) 
K02 = K{\"o}hler \& Petr-Gotzens (2002);  
K11 = Kraus et al. (2011);
L93 = Leinert et al. (1993);
L08 = Loinard et al. (2008);
M03 = Melo et al. (2003); 
H11 = Hu{\'e}lamo et al. (2011);
N12 = Nguyen et al. (2012);
P07 = Prato (2007);
R05 = Ratzka et al. (2005). 
}
\label{t:sample}
\end{deluxetable}

\begin{deluxetable}{lrccccc}
\tablewidth{0pt}
\tablecaption{\emph{Herschel} Observation Log}
\tabletypesize{\footnotesize}
\tablehead{
\colhead{ID}&\colhead{PACS OBs ID}&\colhead{PACS Duration}&\colhead{PACS Date}&\colhead{SPIRE OBs ID}&\colhead{SPIRE Duration}&\colhead{SPIRE Date} \\
\colhead{}&\colhead{}&\colhead{(sec)}&\colhead{(yyyy-mm-dd)}&\colhead{}&\colhead{(sec)}&\colhead{(yyyy-mm-dd)}}
\rotate
\startdata
V1096 Tau    &     1342204863, 1342204864  &   445, 220   &  2010-09-19   & 1342203620   &   445   &  2010-08-24  \\
Hubble 4  &     1342204865, 1342204866 &  445, 220    &  2010-09-19  & 1342203619   &   445   &  2010-08-24 \\
V819 Tau  &      1342216038, 1342216039 &    445, 220 &  2011-03-14  & 1342202256   &   307   &  2010-08-08  \\
HD283572  &      1342204867, 1342204868  & 445, 445   &  2010-09-19   & 1342203618   &   445   &  2010-08-24  \\
L1551-51  &      1342228935  & 895   &  2011-09-19   & 1342203623   &   445   &  2010-08-24  \\
V827 Tau  &     1342228926  &  895  &   2011-09-19   & 1342203622   &   445   &  2010-08-24 \\
RXJ0432.8+1735  &  1342215980    &   670   &   2011-03-13  & 1342203624   &   445   &  2010-08-24  \\
V830 Tau  &   1342202314, 1342202315    &  445, 445   &  2010-08-09   & 1342203621   &   445   &  2010-08-24  \\
RECX 3  &      1342203296 &  670  &  2010-08-22   & 1342202213   &   445   &  2010-08-04  \\
RECX 4  &      1342202897 &  670   &   2010-08-12  & 1342203634   &   445   &  2010-08-24  \\
RECX 6  &   1342209064   & 895   &    2010-11-05 & 1342202214   &   445   &  2010-08-04  \\
RECX 10  &      1342209065    &   895 &   2010-11-05    & 1342203635   &   445   &  2010-08-24  \\
TWA 6  &       1342209479 &    895 &  2010-11-18   & 1342200127   &   445   &  2010-07-09  \\
TWA 7  &    1342212629  &    445 &  2011-01-14T   & 1342200128   &   445   &  2010-07-09  \\
T Cha  &    1342209063  &    670 &    2010-11-05 & 1342203636   &   721   &  2010-08-24  \\
TWA 10  &     1342202559 &  895   &   2010-08-10 & 1342202222   &   445   &  2010-08-05  \\
TWA 17  &     1342203105  &  895  &  2010-08-16    & 1342203563   &   445   &  2010-08-23  \\
Sz 67  &      1342204182, 1342204183   &  445, 445   &  2010-09-09   & 1342203565   &   445   &  2010-08-23  \\
Sz 68  &       1342202365, 1342202366  &   445, 445 &   2010-08-10 & 1342203566   &   307   &  2010-08-23  \\
RXJ1603.2-3239  &   1342204184   &  670   &   2010-09-09  & 1342203568   &   445   &  2010-08-23  \\
Sz 96  &     1342216049, 1342216050 &  220, 220   &  2011-03-14   & 1342203567   &   307   &  2010-08-23  \\
RXJ1622.6-2345  &     1342204193, 1342204194   &   445, 220  &   2010-09-09  & 1342203576   &   445   &  2010-08-23  \\
RXJ1625.2-2455b  &     1342204185, 1342204186   & 445, 445    &    2010-09-09   & 1342203575   &   445   &  2010-08-23  \\
ROXs 3  &   1342215611, 1342215612    &  445, 445   &  2011-03-08  & 1342203574   &   445   &  2010-08-23  \\
ROXR1 13  &      1342216064, 1342216065  &  220, 220   &  2011-03-15   & 1342203074   &   307   &  2010-08-15  \\
RXJ1627.2-2410  &   1342204189, 1342204190 &  445, 220   &   2010-09-09    & 1342203090   &   445   &  2010-08-16  \\
ROXs 31  &    1342204187, 1342204188    &  445, 220   &   2010-09-09   & 1342203572   &   445   &  2010-08-23  \\
RXJ1628.2-2405  &    1342204280, 1342204281    &  445, 220   &  2010-09-10   & 1342203571   &   445   &  2010-08-23 \\
ROXs 39  &        1342215613, 1342215614 &  445, 445  &   2011-03-08  & 1342203073   &   445   &  2010-08-15  \\
ROXs 42C  &   1342216062, 1342216063   & 220, 220    &  2011-03-15    & 1342203570   &   307   &  2010-08-23  \\
ROXs 47A  &     1342216060, 1342216061   &  220, 220   &   2011-03-15  & 1342203569   &   445   &  2010-08-23 
\enddata
\label{obs_ids}
\end{deluxetable}
\begin{deluxetable}{lrrrrrrrrrr}
\tablewidth{0pt}
\tablecaption{\emph{Herschel} Photometry}
\tabletypesize{\tiny}
\tablehead{
\colhead{ID}&\colhead{F$_{70}$}&\colhead{$\sigma$$_{70}$}&\colhead{F$_{160}$}&\colhead{$\sigma$$_{160}$}&\colhead{F$_{250}$}&\colhead{$\sigma$$_{250}$}&\colhead{F$_{350}$}&\colhead{$\sigma$$_{350}$}&\colhead{F$_{500}$}&\colhead{$\sigma$$_{500}$}\\
\colhead{}&\colhead{(mJy)}&\colhead{(mJy)}&\colhead{(mJy)}&\colhead{(mJy)}&\colhead{(mJy)}&\colhead{mJy}&\colhead{(mJy)}&\colhead{(mJy)}&\colhead{(mJy)}&\colhead{(mJy)}}
\startdata
V1096 Tau                     &   $<$ 5.73e+00  &    \nodata  & $<$ 5.33e+01  &  \nodata  & $<$ 2.70e+02  &  \nodata     &   $<$  5.10e+02  &  \nodata       &  $<$  7.50e+02  &  \nodata  \\
Hubble 4                 &   $<$  6.93e+00 &    \nodata  & $<$ 8.72e+01 &  \nodata  &  $<$ 7.50e+02  &  \nodata    &   $<$ 1.05e+03  &  \nodata         &   $<$ 1.20e+03  &  \nodata  \\
V819 Tau                 &        1.54e+01 &    1.03e+00  &     $<$ 2.19e+01 &  \nodata  &  $<$ 1.12e+02  &  \nodata       &   $<$   1.80e+02 &  \nodata        &  $<$ 2.25e+02  &  \nodata  \\
HD283572              &   $<$ 4.67e+00 &    \nodata  & $<$  1.29e+01 &  \nodata  &  $<$ 3.75e+01  &  \nodata     &   $<$  5.25e+01 &  \nodata        &  $<$  6.75e+01  &  \nodata  \\
L1551-51                &  $<$ 2.29e+00&    \nodata  &  $<$ 3.66e+00  &  \nodata  &  $<$ 3.00e+01  &  \nodata     &   $<$   3.75e+01 &  \nodata       &  $<$  5.25e+01  &  \nodata  \\
V827 Tau                 &   $<$ 3.98e+00 &    \nodata  &  $<$ 4.55e+00  &  \nodata  &  $<$ 3.75e+01  &  \nodata     &   $<$   6.00e+01 &  \nodata        &  $<$ 7.50e+01  &  \nodata  \\
RXJ0432.8+1735 &   1.20e+01  &  2.47e+00  & $<$ 1.42e+01 &  \nodata  &   $<$ 2.25e+01  &  \nodata     &   $<$   2.25e+01 &  \nodata        &  $<$ 3.00e+01  &  \nodata  \\ 
V830 Tau                 &  $<$ 1.79e+00 &    \nodata  & $<$ 1.08e+01  &  \nodata  &  $<$ 7.50e+01  &  \nodata     &   $<$   1.05e+02 &  \nodata        &  $<$ 1.20e+02  &  \nodata  \\
RECX 3                  &   6.04e+00  & 1.29e+00  &  $<$ 1.32e+01  &  \nodata  &  $<$ 1.80e+01  &  \nodata        &   $<$   2.25e+01 &  \nodata       &  $<$ 3.00e+01  &  \nodata  \\ 
RECX 4                   &  9.25e+00   & 1.32e+00   & $<$ 6.41e+00  &  \nodata  &  $<$ 2.25e+01  &  \nodata        &   $<$   3.00e+01 &  \nodata      &  $<$ 3.00e+01  &  \nodata  \\ 
RECX 6                    &  $<$ 4.80e+00&    \nodata  &  $<$ 1.65e+01 &  \nodata  & $<$ 2.70e+01  &  \nodata       &   $<$   2.70e+01 &  \nodata         &  $<$ 3.00e+01  &  \nodata  \\
RECX 10                &   $<$  2.98e+00 &    \nodata  &  $<$ 7.51e+00  &  \nodata  &  $<$ 2.25e+01  &  \nodata      &   $<$   2.25e+01 &  \nodata        &  $<$ 2.70e+01  &  \nodata  \\
TWA 6                       &  $<$ 6.03e+00 &    \nodata  &  $<$ 7.67e+00 &  \nodata  &   $<$ 1.65e+01  &  \nodata     &   $<$   1.80e+01 &  \nodata        &  $<$ 2.25e+01  &  \nodata  \\
TWA 7                      &   6.88e+01   & 3.13e+00 &  4.98e+01 &   7.05e+00  &          2.75e+01  &  5.00e+00   &            2.40e+01  &   5.00e+00   &  $<$ 3.00e+01  &  \nodata  \\
T Cha                      & 5.06e+03 & 2.37e+01 & 3.97e+03 & 1.90e+01  &      1.78e+03  &  2.00e+02   &            1.06e+03  &   1.20e+02   &         6.60e+02  &  7.00e+01  \\  
TWA 10                    &   $<$ 3.81e+00 &    \nodata  & $<$ 1.37e+01  &  \nodata  &  $<$ 2.70e+01   &  \nodata      &   $<$   2.70e+01 &  \nodata        &  $<$ 2.25e+01  &  \nodata  \\
TWA 17                  &   $<$ 2.43e+00 &    \nodata    &  $<$ 9.41e+00   &  \nodata  &  $<$ 2.70e+01  &  \nodata       &   $<$   3.00e+01 &  \nodata        &  $<$ 3.75e+01  &  \nodata  \\
Sz 67                        &  $<$ 4.10e+00 &    \nodata  &   $<$ 1.51e+01  &  \nodata  &  $<$ 2.70e+01  &  \nodata      &   $<$   3.00e+01 &  \nodata         &   $<$ 3.00e+01  &  \nodata  \\
Sz 68                        &   2.65e+03   &  5.08e+01 &  $<$ 2.30e+03 &  \nodata  &  $<$ 6.00e+03  &  \nodata      &   $<$   3.00e+03 &  \nodata         &  $<$ 3.00e+03  &  \nodata  \\
RXJ1603.2-3239   &  1.28e+01  &   1.08e+00  &  $<$ 1.44e+01 &  \nodata  &  $<$ 2.25e+01  &  \nodata     &   $<$   3.00e+01 &  \nodata        &  $<$ 3.00e+01  &  \nodata  \\
Sz 96                        & 1.67e+02 & 2.02e+00 & 9.22e+01 &  1.64e+01  &  $<$ 1.80e+02  &  \nodata       &   $<$  3.00e+02 &  \nodata        &  $<$ 4.50e+02  &  \nodata  \\
RXJ1622.6-2345   &   $<$ 5.81e+00  &    \nodata  &  $<$ 2.47e+01  &  \nodata  &   $<$ 7.50e+01  &  \nodata     &   $<$   7.50e+01 &  \nodata        &  $<$ 9.00e+01  &  \nodata  \\
RXJ1625.2-2455b &   2.04e+01  & 1.21e+00 &  8.84e+01 &   1.07e+01  &  $<$ 9.00e+01  &  \nodata     &   $<$   1.20e+02 &  \nodata        &  $<$ 1.35e+02  &  \nodata  \\
ROXs 3                   &    $<$ 3.63e+00 &    \nodata  &  $<$ 3.98e+01 &  \nodata  & $<$ 1.50e+02  &  \nodata        &   $<$   1.50e+02 &  \nodata       &  $<$ 1.50e+02  &  \nodata  \\
ROXR1 13             &    2.15e+03  & 2.58e+02  &  2.31e+03 &   2.91e+02 & $<$ 1.50e+03   &  \nodata       &   $<$   9.00e+02 &  \nodata       &  $<$ 1.00e+03  &  \nodata  \\ 
RXJ1627.2-2410   &   $<$ 7.79e+00 &    \nodata  &  $<$ 1.02e+02  &  \nodata  &  $<$ 1.50e+02  &  \nodata      &   $<$   2.25e+02 &  \nodata       &  $<$ 2.25e+02  &  \nodata  \\
ROXs 31                 &   1.57e+01  &  1.71e+00  &  $<$ 1.15e+02 &  \nodata  &  $<$ 4.50e+02  &  \nodata        &   $<$   5.25e+02 &  \nodata       &  $<$ 1.20e+03  &  \nodata  \\
RXJ1628.2-2405   &    1.03e+02  & 1.70e+00  &   1.59e+02  &   2.75e+01  &  $<$ 3.00e+02  &  \nodata      &   $<$   4.50e+02 &  \nodata       &  $<$ 4.50e+02  &  \nodata  \\   
ROXs 39                 &    $<$ 5.96e+00&    \nodata  &  $<$ 2.05e+01  &  \nodata  &  $<$ 7.50e+01  &  \nodata        &   $<$   9.00e+01 &  \nodata       &  $<$ 1.20e+02  &  \nodata  \\
ROXs 42C               &   2.40e+02  &  3.00e+00  &  7.09e+01  &  9.70e+00& $<$ 1.50e+02  &  \nodata        &   $<$   1.95e+02 &  \nodata       &  $<$ 2.10e+02  &  \nodata  \\ 
ROXs 47A               &   4.63e+01  &  2.75e+00  &   1.48e+02  & 1.47e+01 &  $<$ 2.40e+02  &  \nodata       &   $<$   3.00e+02 &  \nodata        &  $<$ 3.60e+02  &  \nodata  
\enddata
\label{t:photometry}
\tablenotetext{\empty}{NOTE 1: Upper limits are 3$-\sigma$.}
\tablenotetext{\empty}{NOTE 2: The 70 $\mu$m fluxes of ROXR1 13 and ROXs 47A and the 160 $\mu$m fluxes of RXJ1628.2-2405, Sz 96, ROXR1 13, ROXs 42C, and ROXs 47A 
seem to suffer from different levels of contamination by cloud emission and are thus considered to be upper limits.}
\end{deluxetable}
\begin{deluxetable}{lrrrlrrrccccclcrc}
\rotate
\tablewidth{0pt}
\tablecaption{Optical, 2MASS, \emph{Spitzer}, and mm wavelength photometry}
\tabletypesize{\tiny}
\tablehead{
\colhead{ID}&\colhead{V}&\colhead{R$_C$}&\colhead{I$_{C}$}&\colhead{Ref\tablenotemark{1}}&\colhead{J}&\colhead{H}&\colhead{K}&\colhead{F$_{3.6}$}&\colhead{F$_{4.5}$}&\colhead{F$_{5.8}$}&\colhead{F$_{8.0}$}&\colhead{F$_{24}$}&\colhead{Ref\tablenotemark{2}}
&\colhead{$\lambda_{mm}$}&\colhead{F$_{mm}$}&\colhead{Ref\tablenotemark{3}}
\\
\colhead{}   &\colhead{(mag)}& \colhead{(mag)}&\colhead{(mag)}&\colhead{}&\colhead{(mag)}&\colhead{(mag)}&\colhead{(mag)}&\colhead{(mJy)}&\colhead{(mJy)}&\colhead{(mJy)}&\colhead{(mJy)}&\colhead{(mJy)}&\colhead{}&\colhead{(mm)}&\colhead{(mJy)}&\colhead{}    
 }
\startdata
V1096 Tau                 &  12.69   & 11.90 &  \nodata &  N07   &   8.83   & 7.79  &  7.46   & 380 &    242 &      168 &      101 &    11.8 &  P08 &  \nodata &  \nodata &  \nodata \\ 
Hubble 4              & 12.00    & 11.21 & \nodata  & N07    &   8.56   & 7.63  &  7.29  &    426 & 287 & 192 & 109   & \nodata  & P08 &  0.85 & $<$ 9.0 & A05 \\ 
V819 Tau             & 13.20	  & 12.20  &  11.11   &  KH95 &   9.50   & 8.64  &  8.42  & 154 &  92.3 &  68.9  & 39.9 &  21.6  &  P08 & 0.85 & $<$ 9.0 & A05 \\
HD283572           & 9.03	  &    8.56	&    8.57   &  KH95 &  7.41    & 7.00   &  6.86 &  535 &  338 & 229 & 127 & 14.1  & P08 & 0.85 & $<$ 9.0 & A05   \\
L1551-51              & 13.29   &   12.06 &  11.31  &  S89    &    9.69   & 9.05  &  8.85  & 80.5 &   54.4 & 40.0 &  23.5 & 2.44 & E07 &  0.85 &  $<$ 13 & A05 \\ 
V827 Tau               & 12.27	 &   11.39 &  11.28  &     KH95 & 9.16 & 8.49 & 8.23 & 158 &  108 & 71.9  & 39.7 & 4.73 &   L10 &  0.85 &  $<$ 6.0 &  A05 \\ 
RXJ0432.8+1735  & 13.66 & 12.60 & 11.32    &    P06 &  10.00 & 9.23 & 9.02 & 80.5 & 45.0 & 35.4 &  22.4 & 16.9 & E07  &  \nodata &  \nodata &  \nodata \\
V830 Tau               &     12.23  &	 11.37 & 11.37 &   KH95 & 9.32  &  8.61 & 8.42  & 94.4 &  73.4 &  53.0 & 32.4&   3.66 & E07 &   0.85 &  $<$ 6.0  &  A05 \\
RECX 3                 &     14.37 &  13.21 &  11.79 &   S09 & 10.34 &  9.64 & 9.41 &    55.0   &  37.1  &  26.5   &  14.0   &  2.88  & S09  &  \nodata &  \nodata &  \nodata \\
RECX 4                 &     12.73  &  11.80 & 10.81  &  S09  &  9.53 & 8.77  &  8.61 &    117   & 74.8  & 51.5   &  28.7 &     5.80  & S09   &  \nodata &  \nodata &  \nodata\\
RECX 6                 &     14.08  &   13.08   & 11.68 & S09  & 10.23 & 9.58 & 9.29     & 61.4 & 42.3 & 27.8   & 15.5 &   2.01  &  S09  &  \nodata &  \nodata &  \nodata\\
RECX 10               &     12.53   &   11.67    &  10.75  & S09 & 9.65   &  8.92 & 8.73 & 103 & 64.6 & 43.7 & 25.7 & 2.98 &  S09  &  \nodata &  \nodata &  \nodata\\
TWA 6                    &      11.62    & \nodata &   9.94 &   B06 &  8.86  &  8.18 &  8.04 & \nodata & \nodata & \nodata & \nodata &  5.7 & L05 & \nodata & \nodata & \nodata  \\ 
TWA 7                     &     11.65   & \nodata &  9.21  &  B06  &  7.79 &  7.12  & 6.89  & \nodata & \nodata & \nodata & \nodata &  30.2 & L05  &  0.85 & 9.7 & M07 \\
T Cha                      &      11.87  & 11.02       &  10.16     &  A93   &    8.95 & 7.86   &  6.95 & 1490 & 1320 & 1070 & 666 & \nodata & B07 & 1.30 & 105 & L07 \\     
TWA 10                   &     12.96    &  \nodata & 10.49     & B06  &  9.12 &  8.47  &  8.18  & \nodata & \nodata & \nodata & \nodata & 5.10 & L05  &  \nodata &  \nodata &  \nodata\\
TWA 17                    &     12.70  &  \nodata & 10.78    &  B06  & 9.80 &  9.18   &  9.01  & \nodata & \nodata & \nodata & \nodata  &  1.50 & L05  &  \nodata &  \nodata &  \nodata\\
Sz 67                         &   \nodata& 12.85	&    11.49	  &  C07   & 9.99    & 9.36     & 9.12 & 79.0 &   50.0 &   35.0 &    21.2 & 2.68 & E07  &  1.30  &  $<$ 66 &  N97\\
Sz 68                          &   10.36   &  9.60     &    8.86    &   C07   &    7.75   &  6.86  &  6.48 &  1580 & 1420 & 1690 & 2170 & 2860 & E07 & 1.30 & 73 & L07  \\
RXJ1603.2-3239     &    12.71  &  11.90   &   11.05   &  P06   &    9.97   & 9.29  &  9.12  & 66.8 &  44.29 & 34.7 & 20.9 &    9.23 & E07 &  \nodata &  \nodata &  \nodata   \\
Sz 96                         &     13.91  &   13.02   &    11.84  & C07   &    10.12 &  9.34   &    8.95  & 168 & 113 & 138 & 173 &  241  & E07 & 1.30 & $<$ 13 & L07 \\
RXJ1622.6-2345    &     16.04   &   14.45    &   12.98  &   C07   &  11.05 &  10.06 &    9.73     &  43.7 & 32.2 & 22.1 & 13.9 & 2.77 & E07  & 0.85 & $<$ 11 & C08 \\
RXJ1625.2-2455b  &    14.41   &   13.05    &   11.75  &    C07  &  9.58   &   8.58   &    8.26      &  171 & 112 &  81.0 & 48.9 &  5.29 & E07   &  \nodata &  \nodata &  \nodata \\
ROXs 3                      &    13.10  &     12.10   &  11.13   &     C07  &   9.77    &  9.04      &  8.78   &\nodata& 62.5 &\nodata&  26.2 &  3.07 & E07  &  1.30 &  $<$ 30 &  A07 \\
ROXR1 13                 &   14.01   &   12.38    &   10.80   &    C07   &    8.09    &   6.86    &  6.22   &1260 &  878  & 743  & 689  &  1810 & E07 & 0.85 & $<$ 18 & C08 \\
RXJ1627.2-2410     &    18.68  &   18.49    &   15.36  &    C07     &   11.87   &  10.29   &  9.64  &  49.0 &  37.7  & 27.2 & 17.1  &1.93  & E07  &  \nodata &  \nodata &  \nodata  \\
ROXs 31                    &   16.28   &   14.48    &  12.78   &    C07     &   9.99     &  8.72     &   8.12  &   203 & 145 & 106    &62.7   &  7.40   & E07  & 0.85 &  $<$ 22 & A07  \\
RXJ1628.2-2405     &   \nodata&\nodata   &\nodata   &\nodata     &  10.98   &   9.57    &    8.86 & 134 & 129 & 109  &   90.1 & 93.4  &  E07 &  0.85 & $<$ 15 & C08  \\
ROXs 39                    &  12.90    &   11.85    &   10.76   &     C07    &   9.09    &   8.28    &   8.02  &   197 & 138 & 98    &   57.7 & 6.34  & E07   & 1.30 &  $<$ 30 & A07 \\
ROXs 42C                 &  12.05    &  11.06     &  10.13   &      C07    &   8.35    &   7.51    &   7.12  & 575  &  428 &  371 &   397 & 862  & E07 & 0.85 & $<$ 13  & C08\\
ROXs 47A                 &  13.65    & 12.43      &   11.15   &     C07    &    9.24   &   8.35    &   7.92  & 265  &  212 &  152 &  122  & 105  & E07 & 1.30 & $<$ 6.6 & C08  
\enddata
\label{t:photo-rest}
\tablenotetext{1}{References  for optical data are: 
A93= Alcala et al. 1993;
B06 =  Barrado Y Navascues, 2006;
C07 = Cieza et al. 2007;
K95 = Kenyon $\&$ Hartmann, 1995; 
N07 = Norton et al. (2007);
P06 = Padgett et al. 2006; 
S89 = Strom et al. 1989;
S09 = Sicilia-Aguilar et al. 2009.
}
\tablenotetext{2}{References  for \emph{Spitzer}  data are: 
P08 = Padgett et al. (2008), (Taurus delivery document);  
S09 = Sicilia-Aguilar et al. 2009;
B07 = Brown et al. 2007; 
E07 = Evans et al. 2007, (c2d Delivery document);
L05 = Low et al. 2005.
}
\tablenotetext{4}{References  for mm data  are:
A05 = Andrews $\&$ Williams, 2005;
A07 = Andrews $\&$ Williams, 2007
C08 = Cieza et al. 2008a;
M07 = Matthews et al. 2007;
N97 = Nuernberger et al. 1997
L07 = Lommen et al. 2007.
}
\end{deluxetable}
\begin{deluxetable}{lr}
\footnotesize
\tablecaption{Primordial disks, default parameters}
\tablehead{
\colhead{Parameter}&\colhead{Default value\tablenotemark{1}}
}
\startdata
Surface density exponent, $\alpha$     & $-$1 \\
Inclination,  $i$      [deg]                          &  60 \\
Grain size distribution slope, p   &  $-$3.5     \\
Min grain size,  a$_{\mathrm{min}}$  [$\mu$m ] &  0.01     \\
Max grain size,   a$_{\mathrm{max}}$ [$\mu$m]   &  ~3900 \\
Settling exponent, $\xi$    &       0    \\
Outer radius, R$_{\mathrm{out}}$ [AU]        & 100
\label{t:pri-fixed}
\enddata
\tablenotetext{1}{The $\xi$ value is allowed to change, but only if needed to
better fit the silicate feature. Similarly, R$_{out}$ is allowed to change to a lower
value, but only if the target is a close binary system. 
}
\end{deluxetable}

\begin{deluxetable}{lcccccccc}
\tablecaption{Primordial disks, best-fit parameters\tablenotemark{1}}
\tablehead{\colhead{Object}&\colhead{R$_{\mathrm{in}}$}&\colhead{H$_{50}$}&\colhead{$\beta$}&\colhead{M$_{disk}$\tablenotemark{2}}&\colhead{$\xi$}&\colhead{R$_{\mathrm{out}}$}&\colhead{T$_{eff,\star}$\tablenotemark{3}}&\colhead{L$_{\star}$}\\
\colhead{}&\colhead{[AU]}&\colhead{[AU]}&\colhead{}&\colhead{[M$_{\mathrm{JUP}}$]}&\colhead{}&\colhead{[AU]}&\colhead{[K]}&\colhead{[L$_{\odot}$]}
}
\startdata
Sz 68                 &  0.8 & 3.0     & 1.18 & 4.0                &  0.1    & 15    &  4900 & 6.7    \\
Sz 96                 &  0.1 & 6.0     & 1.07 & 0.9                &  0.2    & default & 3580   &  0.9 \\
RXJ1628.2-2405        & 0.8  &0.17 & 1.15   & 7$\times$10$^{-2}$ & default & default & 4350  &   0.9 \\
ROXs 42C (inner disk) & 0.09 & 1.0 & 1      & 2$\times$10$^{-4}$ & default &  3  &  4200  & 2.7  \\
ROXs 42C (outer disk) & 3.0  & 4.5 & 1      & 8$\times$10$^{-3}$ & default &  30 &  4200 & 2.7 \\
ROXs 47A (inner disk) & 0.05 & 5 &  1.1      & 2$\times$10$^{-5}$  & default &  1.5 & 4350 & 1.4 \\
ROXs 47A (outer disk) & 1.5  & 5  & 1.1      & 2$\times$10$^{-3}$ & default &  default & 4350 & 1.4 \\
ROXR1 13              &  \nodata  &  \nodata &  \nodata &  \nodata &  \nodata & \nodata &  \nodata &  \nodata
\label{t:pri-fit}
\enddata
\tablenotetext{1}{Default values are listed in Table 5.}
\tablenotetext{2}{M$_{disk}$ is not formally a free parameter, but
is calculated by integrating $\Sigma$ over radius and assuming a gas
to dust mass ratio of 100. For convenience, we report M$_{disk}$ instead
of the corresponding free parameter $\Sigma_{50}$.
See Section~\ref{prim-models} regarding
important caveats on the M$_{disk}$ parameter.}
\tablenotetext{3}{T$_{eff,\star}$ and L$_{\star}$ are fixed model parameters.
They were derived from the spectral types (using the temperature scale given by Kenyon \& Hartmann 
et al. 1995) and applying a bolometric correction (from Hartigan et al. 1994) to the extinction-corrected
J-band fluxes.}
\end{deluxetable}
\begin{table}
\caption{Best fit results for the debris disk sample\label{table:results}}
\begin{center}
\begin{tabular}{lcccccc}
\hline \hline
Star & $r_{\mathrm{in}}$ & $r_{\mathrm{out}}$ & $M_{\mathrm{dust}}$ & L$_{disk}$/L$_{\star}$\tablenotemark{1} &  T$_{eff,\star}$\tablenotemark{2} & L$_{\star}$ \\
 & $[$AU$]$ & $[$AU$]$ & $[M_{\oplus}]$ &  & [K] & [L$_{\odot}$] \\
\hline
Warm debris disks & & & & && \\
\hline
V819\,Tau           & 1.1 & 36   & 3.8\,$\times 10^{-3}$ & 10$^{-3.7}$ & 4060 & 0.9  \\
RXJ0432.8+1735 & 1.2 & 13     & 1.5\,$\times 10^{-3}$ & 10$^{-3.4}$  & 3580  &  0.4 \\
RECX\,3               & 4.3 & 200 & 4.0\,$\times 10^{-2}$ & 10$^{-4.4}$ & 3470  & 0.13\\
RECX\,4               & 2.5 & 100   & 8.3\,$\times 10^{-3}$ & 10$^{-4.4}$ & 3720 &  0.3 \\
TWA7                   & 3.3 & 423 & 1.6\,$\times 10^{-1}$ &  10$^{-4.3}$ & 3580  & 0.3  \\
 RXJ1603.2-3239 & 13 & 15   & 3.5\,$\times 10^{-3}$ &  10$^{-4.0}$ &  4060 &  0.5\\
\hline
Cold debris disks & & & & && \\
\hline 
RXJ1625.2-2455b\tablenotemark{3}       &  600 & 886 & 12 & 10$^{-4.7}$  &   3850 &  0.9 \\
ROXs\,31              &  40 &  357 & 3.2\,$\times 10^{-1}$ & 10$^{-4.3}$  & 4060 & 1.4 \\
\hline
\end{tabular}
\end{center}
\tablenotetext{1}{calculated by integrating the stellar photosphere and disk model over frequency}.
\tablenotetext{2}{T$_{eff,\star}$ and L$_{\star}$ are fixed model parameters. They were 
derived from the spectral types (using the temperature scale given by Kenyon \& Hartmann et al. 1995) 
and applying a bolometric correction (from Hartigan et al. 1994) to the extinction-corrected 
J-band fluxes.}
\tablenotetext{3}{Even though this object was detected as a point source at both 70 and 160 $\mu$m,
its 160 $\mu$m image shows significant extended emission north-west of the source
(see Figure~2), which could be affecting our photometry.}  
\end{table}

\end{document}